\documentclass[journal]{IEEEtran}
\ifCLASSINFOpdf
\else
\fi
\usepackage{graphicx} 
\usepackage{epsfig} 
\usepackage{amsmath} 
\usepackage{amssymb}  
\usepackage{amsthm}
\usepackage{bm}
\usepackage{subfig}
\usepackage{url}
\usepackage{color}
\usepackage{cite}
\usepackage{multirow}
\usepackage{diagbox}

\usepackage[ruled]{algorithm2e}

\SetCommentSty{mycommfont}
\usepackage{algpseudocode}
\algnewcommand\algorithmicforeach{\textbf{for each}}
\algdef{S}[FOR]{ForEach}[1]{\algorithmicforeach\ #1\ \algorithmicdo}
\usepackage{newtxmath}
\usepackage{subfig}

\newtheorem{theorem}{Theorem}

\begin{document}
	
	\title{Uncertainty Quantification of Hyperspectral Image Denoising Frameworks based on Sliding-Window Low-Rank Matrix Approximation}
	
	\author{Jingwei~Song,
		Shaobo~Xia,
		Jun~Wang,
		Mitesh~Patel,
		Dong~Chen,~\IEEEmembership{Member,~IEEE}
		\thanks{This work was supported in part by the National Natural Science Foundation of China under Grant 41971415; in part by the Natural Science Foundation of Jiangsu Province under Grant BK20201387; in part by the Toyota Research Institute; and in part by the Talent Startup Project of Zhejiang A \& F University Scientific Research Development Foundation under Grant 2034020104. (Corresponding author: Dong Chen)}
		\thanks{Jingwei Song is with the College of Civil Engineering, Nanjing Forestry University, Nanjing 210037, China, and the Department of Naval Architecture and Marine Engineering, University of Michigan, Ann Arbor, MI 48109, USA.
			e-mail:(jingweso@umich.edu); this work was partly performed during the tenure of his position at the FXPAL inc.}
		\thanks{Shaobo Xia is with the College of Environment and Resources, Zhejiang A\&F University, Hangzhou, Zhejiang, China, and also with the Department of Geomatics Engineering, University of Calgary, T2N 1N4, Canada. e-mail: (shaobo.xia@zafu.edu.cn). }
		\thanks{Jun Wang is with the Institute of Remote Sensing and Digital Earth, Chinese Academy of Sciences, University of Chinese
			Academy of Sciences, Beijing, China. 
			e-mail:(wangjun@radi.ac.cn). }
		\thanks{Mitesh Patel is with FX  Palo  Alto  Laboratory  Inc.  Palo  Alto,  CA  -  94304,USA. 
			e-mail:(patelmiteshn@gmail.com). }
		\thanks{Dong Chen is with the College of Civil Engineering, Nanjing Forestry University, Nanjing 210037, China. e-mail: (chendong@njfu.edu.cn).}
	}
	
	\markboth{IEEE Transactions on Geoscience and Remote Sensing}%
	{Shell \MakeLowercase{\textit{et al.}}: Bare Demo of IEEEtran.cls for Journals}
	
	\maketitle
	
	\begin{abstract}
		Sliding-window based low-rank matrix approximation (LRMA) is a technique widely used in hyperspectral images (HSIs) denoising or completion. However, the uncertainty quantification of the restored HSI has not been addressed to date. Accurate uncertainty quantification of the denoised HSI facilitates applications such as multi-source or multi-scale data fusion, data assimilation, and product uncertainty quantification since these applications require an accurate approach to describe the statistical distributions of the input data. Therefore, we propose a prior-free closed-form element-wise uncertainty quantification method for LRMA-based HSI restoration. Our closed-form algorithm overcomes the difficulty of handling uncertainty in HSI patch mixing caused by the sliding-window strategy used in the conventional LRMA process. The proposed approach only requires the uncertainty of the observed HSI and provides the uncertainty result relatively rapidly and with similar computational complexity as the LRMA technique. We conduct extensive experiments to validate the estimation accuracy of the proposed closed-form uncertainty approach. The method is robust to at least $10\%$ random impulse noise at the cost of $10-20\%$ of additional processing time compared to the LRMA. The experiments indicate that the proposed closed-form uncertainty quantification method is more applicable to real-world applications than the baseline Monte Carlo test, which is computationally expensive. 
	\end{abstract}
	
	\section{INTRODUCTION}
	\label{Section:Intro}
	Data uncertainty, which refers to the lack of sureness about the data \cite{dehghan2006measurement}, is an essential factor to consider when utilizing data and derived products \cite{foody2003uncertainty}. When the data uncertainty is known, users can determine whether a product is suitable for specific applications or can balance the contributions of different data sources. For example, in \cite{verhoef2007coupled}, the minimum detectable deformation can be estimated based on the uncertainty measures in a point cloud so that engineers can develop appropriate project plans, including sensor selection and set-up schemes \cite{chen2017determination}. In multiple data fusion tasks, the label uncertainties of hyperspectral images (HSIs) and point clouds are often considered to achieve high fusion accuracy \cite{du2019multiresolution}.
	\par
	
	In theory, uncertainty can be classified as aleatoric uncertainty, and epistemic uncertainty \cite{senge2014reliable}. Aleatoric uncertainty is caused by noise in the data and the processing steps. In contrast, epistemic uncertainty is introduced into the observation models, and data post-processing steps \cite{foody2003uncertainty,oxley2017uncertainties,chen2017determination,bastin2002visualizing,dehghan2006measurement}. In the absence of ground truth data to measure the model performance, the epistemic uncertainty is ignored, and the output uncertainty is defined as the aleatoric uncertainty due to the propagation of an arbitrary model. Most studies have defined the probabilistic density function of the noise as a mixture of the normal distribution and the random distribution. Assuming that random noise is significantly low, the uncertainty can be described by the Gaussian distribution. In HSI restoration, symmetric uncertainty based on mutual information was proposed in \cite{sarhrouni2012application} and was used in HSI classifications. In HSI unmixing, the end member uncertainty of unmixing methods based on the normal compositional model is defined as the covariance matrix, which is used to predict the unmixing errors without knowing the ground truth \cite{zhou2016spatial}. Uncertainty has been described differently in different studies. \cite{dehghan2006measurement} proposed an entropy-based uncertainty measure for evaluating the classification results of multi-spectral remote sensing images. The calculated entropy, which was defined as the pixel-wise classification uncertainty, had a linear relation with the accuracy of the classification results.  In \cite{polo2012estimating}, the uncertainty of 3D point clouds was defined as the measurement repeatability, which was represented by the median and the two percentiles (5\% and 95\%) related to repeated measurements. \par
	
	This paper focuses on the uncertainty quantification in HSI denoising. HSIs consisting of hundreds or thousands of spectral bands have been widely used in various applications (e.g., land cover mapping, pollution monitoring, and crop yield estimation) due to their rich spectral information compared to the spatial information. In HSI data processing, image denoising is regarded as an important preprocessing step for advanced applications and has been researched intensively in the past decades \cite{ghamisi2017advances}. According to \cite{zhang2019hybrid}, existing HSI denoising methods can be categorized into three groups: filter-based methods, model optimization methods, and deep learning methods. Filter-based methods try to remove noise using various filters. Classical methods in this category are based on nonlocal filters \cite{buades2005non,maggioni2012nonlocal,qian2012hyperspectral,dabov2007image}. Model optimization methods can be further divided into total variation methods \cite{yuan2012hyperspectral}, sparse representation-based \cite{li2016noise}, and sliding-window low-rank matrix-based methods \cite{zhang2013hyperspectral}. The major advantages of this group of denoising methods are utilizing image priors and considering spatial and spectral features simultaneously. These methods are often combined to achieve better performance \cite{zhuang2018fast,he2015hyperspectral}. The last group is based on deep learning frameworks, which have attracted increasing attention in the past three years \cite{yuan2018hyperspectral}. A recent deep learning-based framework was proposed in \cite{zhang2019hybrid} to remove hybrid noise in a spatial-spectral framework. \par
	
	HSI denoising based on the sliding-window low-rank matrix approximation (LRMA) framework has become a state-of-the-art technique in the community \cite{zhang2013hyperspectral,he2015total}. These methods are model optimization methods aimed at exploiting the low-dimensional structure in the high-dimensional data space. LRMA-based studies \cite{bioucas2008hyperspectral,zhang2013hyperspectral,zhuang2018fast} assume that the noise-free HSI can be represented by a low-rank matrix. Thus, HSI denoising can be regarded as a classical sliding-window low-rank matrix recovery problem. A representative HSI denoising framework based on LRMA is described in \cite{zhang2013hyperspectral}, where a clean image, impulse noise, and Gaussian noise were formulated separately in the objective function, which was minimized under a low-rank assumption. Many variants of LRMA-based HSI denoising methods have been proposed in the past five years \cite{he2015total,chen2017denoising,zhuang2018fast,fan2018spatial,chen2019hyperspectral}. For example, \cite{he2015total} proposed an LRMA-based framework containing a total-variation regularization term to preserve spatial information. The spatial-spectral structure in HSI denoising was also exploited by replacing the sliding-window low-rank matrix analysis with low-rank tensor approximation \cite{chen2019hyperspectral}. In general, LRMA-based HSI denoising methods are state-of-the-art frameworks that have been widely used in this area.\par
	
	Despite the progress in LRMA-based HSI denoising, no studies investigated the uncertainty quantification of denoised HSI to date. Some works aimed to determine the noise of the raw observations \cite{aiazzi2006noise,acito2011signal}, but the uncertainty of denoised HSI has not yet been addressed, although it is critical in subsequent applications, as discussed above. However, the quantification of uncertainties of denoised HSIs based on LRMA is difficult due to the nonlinear discontinuous functions in LRMA models (e.g., nuclear norm), which hinders the deduction of the closed-form uncertainty propagation formulations. Recently, \cite{chen2019inference} proposed an optimal uncertainty quantification and inference method for noisy matrix completion. Inspired by this work, we expect that it is possible to quantify the uncertainties of outputs from LRMA-based HSI denoising frameworks.\par

	A sliding-window strategy is always used in HSI denoising to handle the extensive data volume of HSI \cite{he2015total,zhang2013hyperspectral}; the denoised element values are obtained by averaging the values from several traversed patches. The mixed values from multiple patches pose difficulty in quantifying the uncertainty of the estimations. Unlike the conventional perspective of multiple observations, which obey the same independent and identical distribution (i.i.d.), multiple overlapping sliding windows are inter-connected statistically. Thus, the covariance of the overlapping window should also be addressed. To tackle these problems, we introduce and improve the method described in \cite{chen2020noisy} to quantify the uncertainty of the low-rank-based HSI approach. Note that we assume the input HSI is a mixture of Gaussian noise and random noise. Previous studies \cite{acito2011signal,qian2012hyperspectral,zhuang2018fast} already proved that the Poisson noise was the primary concern in real HSI data but could be approximated by additive Gaussian noise. Based on these results, Gaussian noise with mean and covariance is suitable to establish the uncertainty propagation model.\par 
	
	This research does not aim at providing a new LRMA formulation. Instead, we propose a general uncertainty quantification method for the existing LRMA approach. The main contributions of this paper are threefold,
	\begin{enumerate}
		\item To the best of our knowledge, the uncertainty quantification of the denoised HSI based on LRMA is presented for the first time in this paper. We provide a closed-form time-efficient uncertainty propagation model to predict the element-wise uncertainty of the denoised HSI without any ground truth data.
		\item To solve the ``uncertainty mixing'' problem caused by the sliding-window strategy, we deduce a weighted average uncertainty formulation, which proves effective based on extensive experiments. 
		\item The proposed uncertainty estimation approach is independent of the choice of the LRMA algorithm. It is based on the global minimum assumption of the LRMA solver; thus, a more accurate solver leads to better uncertainty description.
	\end{enumerate}
	
	\par The rest of this paper is organized as follows. In Section \ref{Section_method}, we briefly summarize the canonical formulation of LRMA-based HSI denoising. Section \ref{Section_uncertianty_quantify} presents the proposed algorithm for the uncertainty propagation in a classical LRMA-based HSI denoising framework. Section \ref{Section_results} shows experiments on both simulated and real HSI data. The discussion of the parameter settings and limitations is also presented in Section \ref{Section_results}. Section \ref{Section_Con} summarizes this paper.

	\begin{figure*}[!h]
		\centering
		\centering
		\includegraphics[width=1\linewidth]{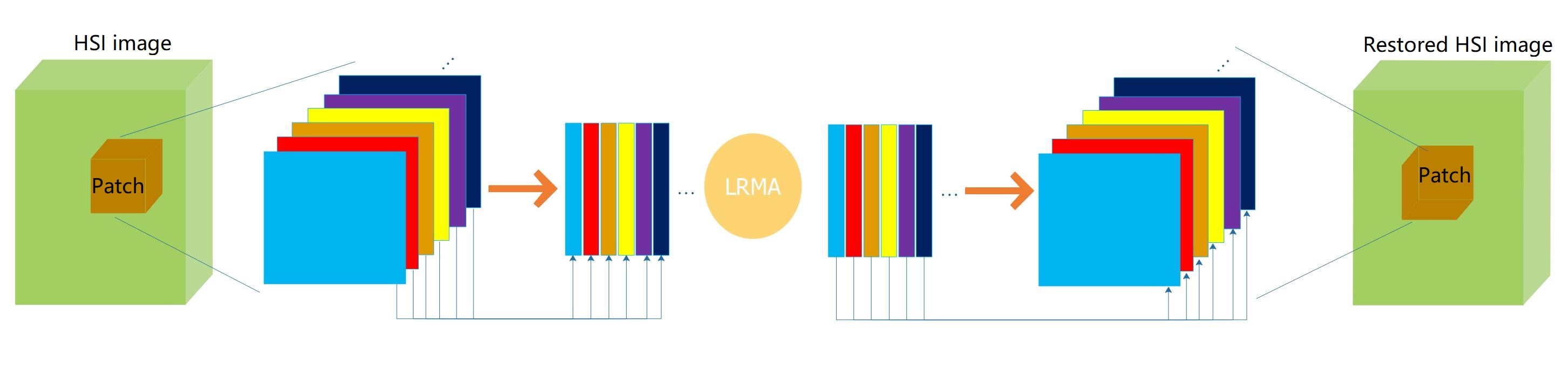}
		\caption{Illustration of the routine process of HSI denoising. The 3D matrices at each end represent the raw and denoised HSI, respectively. The sliding window operates on each patch of the matrix. The selected 3D patch matrix is converted to a 2D matrix by concatenating each band of the 2D matrix as a 1D vector. The permutated 2D matrix goes through the LRMA process to generate the denoised matrix. Symmetrically, the denoised 2D matrix is inversely permutated to the 3D patch matrix and constitutes the final restored HSI in 3D.}
		\label{fig:diagram_hyperspectral_LRMA}
	\end{figure*}

	\section{LRMA algorithm for LRMA HSI denoising} 
	\label{Section_method}
	
	We cite \cite{zhang2013hyperspectral} as a typical LRMA approach and discuss the uncertainty evaluation based on the LRMA approach. Section \ref{Section_method} is a brief formulation of the routine LRMA method \cite{zhang2013hyperspectral}. The results provided in Section \ref{Section_method} have been strictly tested and validated in different LRMA studies. Section \ref{Section_uncertianty_quantify} describes the proposed uncertainty estimation method for an LRMA approach, as described in \cite{zhang2013hyperspectral}. We emphasize that this work DOES NOT aim at proposing another LRMA formulation. The goal of this approach is to quantify the uncertainty of LRMA denoised images in a closed-form manner.\par

	\subsection{Revisiting the general HSI restoration process}
	\label{section_2A}
	Fig. \ref{fig:diagram_hyperspectral_LRMA} demonstrates the routine process of the LMRA-based HSI restoration. The HSI is processed with a 3D sliding window to reduce the computational burden. For each 3D patch in the window, all the bands are vectorized and stacked into the 2D matrix. The permutated 2D matrix goes through the routine LMRA process, and the result is re-permutated to recover the 3D denoised patch. Finally, all the patches are summarized and averaged to create the restored HSI.\par
	
	An HSI acquired by a sensor is typically modeled as a 3D image composed of Gaussian noise and random noise:
	\begin{equation}
	\label{Eq_observation_noise}
	\mathbf{W}=\mathbf{W}^\star+\mathbf{G}+\mathbf{N},
	\end{equation}
	where $\mathbf{W} \in \mathbb{R}^{M \times N \times P}$ is the image acquired from the hyperspectral sensor with the size $M \times N$ and $P$ channels, $\mathbf{W}^\star \in \mathbb{R}^{M \times N \times P}$ is the noise-free image denoted as the ground truth, $\mathbf{G} \in \mathbb{R}^{M \times N \times P}$ is the element-wise i.i.d. Gaussian noise and $\mathbf{N} \in \mathbb{R}^{M \times N \times P}$ is the sparse and random noise (impulse signal). The random noise $\mathbf{N}$ is normally considered negligible with regard to $\mathbf{G}$, and the data quality quantification process is inferring the statistical distribution of $\mathbf{G}$. In practice, the ground truth HSI $\mathbf{W}^\star$ cannot be obtained. Thus, we denote $\hat{\mathbf{W}}$ as the estimated denoised HSI calculated from studies based on the low-rank assumption \cite{zhuang2018fast,he2015hyperspectral}. \par
	
	The goal of HSI denoising is to recover the denoised $\hat{\mathbf{W}}$ with the 3D HSI noisy observation $\mathbf{W}$ and the exact rank prior. {Assuming that the random noise $\mathbf{N}$ is negligible and the residual follows the i.i.d. zero-mean Gaussian distribution, the restored image $\hat{\mathbf{W}}$ is recovered with the following formulation:}
	
	\begin{equation}
	\label{Eq_image_recovery0}
	\begin{array}{l}{\underset{\hat{\mathbf{W}}}{\min}\  \mu \operatorname{rank}(\hat{\mathbf{W}}^\sharp)+\|\mathbf{G}\|^2_\mathrm{F}, \text {\  s.t. }}  {\hat{\mathbf{W}}=\mathbf{W}+\mathbf{G}}\\
	\text {\ such that }
	{\hat{\mathbf{W}}^\sharp=g(\hat{\mathbf{W}})}
	\end{array},
	\end{equation}
	
	\noindent where $\mu$ is the hyperparameter balancing the two terms. $\operatorname{rank}(\cdot)$ constraints the rank of the input matrix to be no greater than the given rank. {$||\cdot||_\mathrm{F}$ is the Frobenius norm that minimizes the residuals in Gaussian distribution.} $g(\cdot)$ permutates the 3D HSI to a 2D matrix by vectorizing and concatenating to a general 2D matrix (Fig. \ref{fig:diagram_hyperspectral_LRMA}). The low-rank formulation was originally proposed by \cite{wright2009robust} and was called ``robust principal component analysis” (RPCA). The work in \cite{zhang2013hyperspectral} introduced RPCA into LRMA scenario.\par
	
	\subsection{The sliding-window in the HSI restoration process}
	\label{section_2B}
	HSI restoration differs greatly from conventional image processing due to the data size. The extremely large permutated 2D matrix hinders the iterative regular singular value decomposition (SVD) process, which is essential in solving Eq. (\ref{Eq_image_recovery0}). The HSI is segmented into small patches (denoted as the overlapped sliding window) and solved in parallel to overcome the heavy computational burden in minimizing the objective function due to the huge HSI size. Previous studies regarded the sliding-window strategy as an engineering compromise and ignored the formal mathematical presentation. However, we realize that the unique sliding-window structure significantly impairs the uncertainty quantification. {By following  \cite{zhang2013hyperspectral}, we obtain a sliding-window version that minimizes all patches in Eq. (\ref{Eq_image_recovery0}):}
	\begin{equation}
	\begin{aligned}
	\label{Eq_image_recovery}
	&\underset{\hat{\mathbf{W}}^{(ijk)}}{\min} 
	& & \mu\operatorname{rank}(\hat{\mathbf{W}}^{\sharp{(ijk)}})+\frac{1}{2}\|\mathbf{W}^{(ijk)}-\hat{\mathbf{W}}^{(ijk)}\|_\mathrm{F}^{2}\\
	&\text {such that}
	& &{\hat{\mathbf{W}}^{\sharp{(ijk)}}=g(\hat{\mathbf{W}}^{(ijk)})}\\
	&&&\text { }(i, j, k) \in \Omega\\
	\end{aligned},
	\end{equation}
	\noindent where the matrices $\mathbf{W}^{(ijk)}$, $\hat{\mathbf{W}}^{(ijk)}$ and $\hat{\mathbf{W}}^{\sharp{(ijk)}}\in \mathbb{R}^{J \times J}$ are the patch indexed in ${(ijk)}$ with regard to $\mathbf{W}$, $\hat{\mathbf{W}}$ and $\hat{\mathbf{W}}^\sharp$. $J$ is the size of the permutated matrix. $\Omega \subseteq\{1, \cdots, M'\} \times\{1, \cdots, N'\} \times\{1, \cdots, P'\}$ is the number of patches in the HSI. After all the patches are optimized, the optimal $\hat{\mathbf{W}}$ is restored by averaging the overlapped patches $\hat{\mathbf{W}}^{(ijk)}$ element-wisely. By ignoring the random noise, we obtain the observation corrupted with the element-wise Gaussian noise:
	
	\begin{equation}
	\begin{aligned}
	\label{Eq_Wsharp_noise}
	&\mathbf{W}^{(i j k)}\approx{\mathbf{W}}^{\star{(i j k)}}+\mathbf{G}^{(i j k)}, \quad \mathbf{G}^{(i j k)} \stackrel{\mathrm{i.i.d.}}{\sim} \mathcal{N}(0, \sigma_0^{2}), \\
	\end{aligned}
	\end{equation}
	\noindent where $\sigma_0^{2}$ is the variance of the noise obtained from the image or estimated by algorithms such as described in \cite{bioucas2008hyperspectral}. $\mathbf{G}^{(i j k)}$ is the Gaussian noise in patch $(i j k)$. In previous studies, the choice of rank was often estimated by manual-tuning strategies based on the differences between the reconstructed signal and the original one \cite{zhuang2018fast}.   \par
	
	The idea of Eq. (\ref{Eq_image_recovery}) is to convert the 3D patch into a typical 2D small-size matrix so that the conventional low-rank factorization algorithm can be executed with limited computational resources. The equation is often solved with a Lagrange multiplier or an augmented Lagrange multiplier (ALM); off-the-shelf low-rank solvers can be used. In this work, we choose the GoDec algorithm proposed by \cite{zhou2011godec} to be consistent with the LRMA method \cite{zhang2013hyperspectral}. However, it should be emphasized that the proposed uncertainty quantification method is independent of the solver since it only assumes that the estimation of the LRMA is close to the minimum. The better the performance of the LRMA solver, the more accurate the confidence interval of our method describes. \par
	
	After all small overlapped patches have gone through the process of Eq. (\ref{Eq_image_recovery}), the optimized denoised patches are summed to restore the HSI:
	
	\begin{equation}
	\label{Eq_image_stacking}
	\hat{\mathbf{W}} = \left[\sum_{i,j,k \in \Omega} \mathit{f}_p ({\hat{\mathbf{W}}^{(ijk)}})\right] \oslash \mathbf{Q},
	\end{equation}
	\noindent where $\mathit{f}_p(\cdot)$ is the padding function to convert the patch to the same size of $\hat{\mathbf{W}}$ by padding the rest of the elements with 0. The matrix $\mathbf{Q}\in \mathbb{R}^{M \times N \times P}$ counts the number of the patches valid on each voxel. $\oslash$ is the Hadamard division operator.\par

	\section{Uncertainty quantification method for LRMA-based HSI denoising} 
	\label{Section_uncertianty_quantify}
	Based on the above-mentioned formulation, we propose a closed-form element-wise uncertainty quantification method that can handle the sliding-window LRMA-based HSI denoising.\par

	\subsection{The general framework}
	\label{section_2C}
	
	\begin{figure}[!h]
		\centering
		\centering
		\includegraphics[width=1\linewidth]{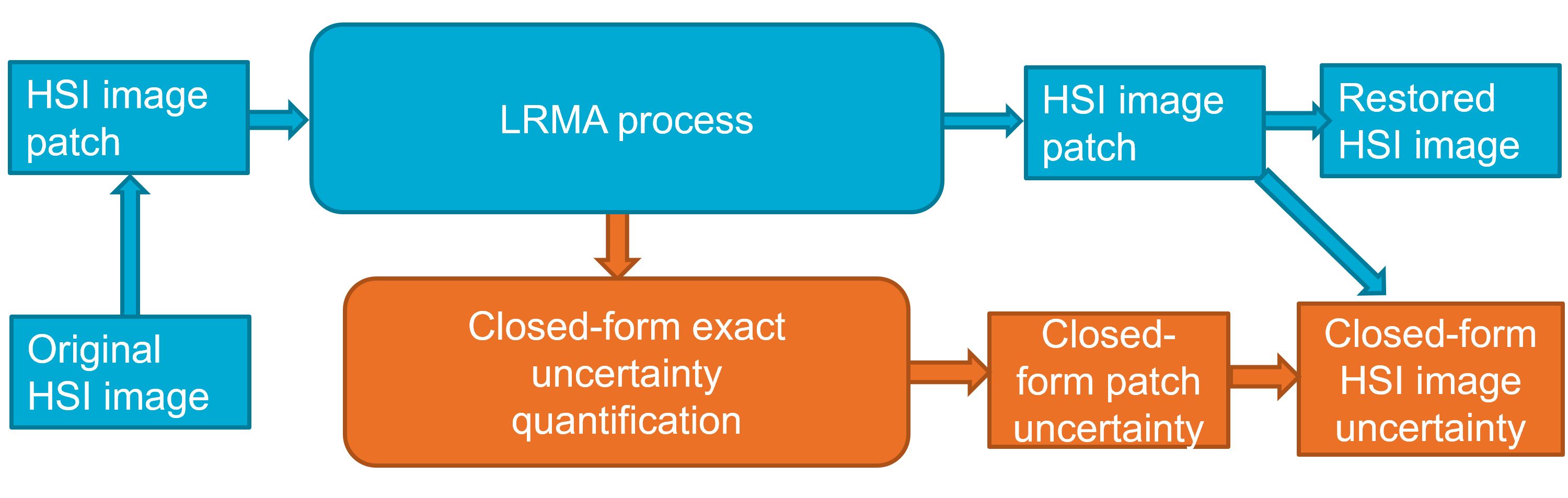}
		\caption{The general framework of the proposed approach. The blue boxes represent the typical LRMA process (\cite{zhang2013hyperspectral} as an example). In addition to the LRMA HSI denoising, our work aims at providing an extra module to quantify the uncertainty of the output and the restored HSI using the LRMA algorithm.}
		\label{fig:framework_uncertainty_RS}
	\end{figure}
	
	Fig. \ref{fig:framework_uncertainty_RS} illustrates the general framework and the relationship with the routine LRMA approach. In the LRMA process, the temporary variable, i.e., the 2D sliding-window patch, is utilized to provide the uncertainty to the HSI patch based on our closed-form formulation. After the HSI patch and its associated closed-form uncertainty have been generated, the routine LRMA approach averages them to yield the restored HSI. We also provide a closed-form uncertainty propagation to quantify the final restored HSI. As shown in Fig. \ref{fig:framework_uncertainty_RS}, in all processes, the proposed approach is independent of the choice of the low-rank solver. The method can also be easily integrated into the derivations of the LRMA process. In addition, since our method is a closed-form solution and only utilizes temporary variables, the processing time is relatively short compared to the routine LRMA process. \par 
	
	One major contribution of this work is that it provides an element-wise closed-form solution to quantify the uncertainty in a time-efficient manner. For the restored HSI $\hat{\mathbf{W}}$, one straightforward approach of the element-wise uncertainty quantification is conducting the Monte Carlo test with the given statistical distribution of the observation $\mathbf{W}$. However, the size of HSIs is notoriously large, resulting in a heavy computational burden to conduct numerous Monte Carlo tests. Roughly, a one-time process in \cite{zhang2013hyperspectral} on a $100\times100\times100$ HSI requires 20 minutes with a typical commercial desktop computer. Since a robust uncertainty quantification requires a reasonable number of trials, the Monte Carlo test is inefficient for use in the element-wise HSI uncertainty quantification.\par
	
	Another benefit of this research is that it solves the uncertainty propagation of the sliding-window process. The uniqueness of HSIs is that the permutated 2D matrix after dimensionality reduction is still relatively large. The solution cannot be achieved with a one-time LRMA process. Instead, researchers have adopted the sliding-window strategy to segment the HSI into many small patches and process them parallel. The different statistical properties of each overlapped patch make it challenging to quantify the uncertainty of the final restored HSI. We solve this problem and present a closed-form solution in the proposed sliding-window element-wise closed-form HSI uncertainty quantification.  \par

	\subsection{Uncertainty quantification for the exact low-rank factorization}
	\label{section_exact_structure}
	
	This section focuses on solving the uncertainty quantification of the patch-wise low-rank factorization process (Eq. (\ref{Eq_image_recovery})). We follow the example of \cite{chen2019inference} and draw a similar conclusion for the uncertainty quantification of the exact low-rank patch factorization. A concise and clear proof is also provided. HSI patch-based averaging (Eq. (\ref{Eq_image_stacking})) will be discussed in the next subsection. A recent study \cite{chen2019inference} proposed an uncertainty quantification method for the exact low-rank 2D matrix. This element-wise algorithm can quantify the confidence interval for a common low-rank based 2D matrix denoising algorithm. Synthetic and real-world tests validate the efficiency of this method. We introduce this approach and modify it to enable the quantification of the uncertainty of each estimated 3D HSI patch $\hat{\mathbf{W}}^{(ijk)}$ using Eq. (\ref{Eq_image_recovery}). We also provide concise proof different from \cite{chen2019inference}.\par

	In the patch optimization process shown in Eq. (\ref{Eq_image_recovery}), we first aim to quantify the uncertainty of the temporary 2D square matrix $\hat{\mathbf{W}}^{\sharp{(i j k)}}$ ($\hat{\mathbf{W}}^{\sharp{(i j k)}} =g(\hat{\mathbf{W}}^{{(i j k)}})$ and it can be any patch in $\Omega$). We denote the ground truth of $\hat{\mathbf{W}}^{\sharp{(i j k)}}$ and $\hat{\mathbf{W}}^{{(i j k)}}$ as ${\mathbf{W}}^{\sharp\star{(ijk)}}$ and ${\mathbf{W}}^{\star{(i j k)}}$. For the sake of conciseness, the rank-r SVD decomposition is ${\mathbf{W}}^{\sharp\star{(i j k)}}=\mathbf{U}^{\star(i j k)} \Sigma^{\star(i j k)} \mathbf{V}^{\star(i j k) \top}$. For conciseness, we simplify $(\mathbf{U}^{\star(i j k)}, \Sigma^{\star(i j k)}, \mathbf{V}^{\star(i j k)})$ by ignoring the patch index $(ijk)$ and rewrite it as $(\mathbf{U}^{\star}, \Sigma^{\star}, \mathbf{V}^{\star})$. Two matrices are defined as the product of the unitary matrix and the squared singular matrix is: $\mathbf{X}^{\star} \triangleq \mathbf{U}^{\star} \Sigma^{\star 1 / 2} \in \mathbb{R}^{K\times r}$, $\mathbf{Y}^{\star} \triangleq \mathbf{V}^{\star} \Sigma^{\star 1 / 2} \in \mathbb{R}^{L \times r}$. The following rules apply:\par

	\begin{equation}
	\begin{aligned}
	\mathbf{X}^{\star \top} \mathbf{X}^{\star}=\mathbf{Y}^{\star \top} \mathbf{Y}^{\star}=\Sigma^{\star}.
	\end{aligned}
	\end{equation}

	Following the above definition, the ground truth 2D temporary matrix is $\mathbf{W}^{\sharp\star{(i j k)}} = \mathbf{X}^{\star} \mathbf{Y}^{\star \top}$. Similarly, we define the estimated version $\hat{\mathbf{W}}^{\sharp{(i j k)}} \approx \hat{\mathbf{X}} \hat{\mathbf{Y}}^\top$ where $\hat{\mathbf{X}}$ and $\hat{\mathbf{Y}}$ are the estimated matrices regarding $\mathbf{X}^{\star}$ and $\mathbf{Y}^{\star}$. We also obtain the estimated version of $\Sigma^{\star}$ as $\hat{\Sigma}$:\par
	
	\begin{equation}
	\label{Eq_XtX_Sigma}
	\begin{aligned}
	\hat{\Sigma}=\hat{\mathbf{X}}^{\top} \hat{\mathbf{X}}=\hat{\mathbf{Y}}^{\top} \hat{\mathbf{Y}}\approx\Sigma^{\star}.
	\end{aligned}
	\end{equation}
	
	{It should be noticed that the factorization of $\hat{\mathbf{W}}$ is not unique. For each pair of estimated auxiliary matrices ($\hat{\mathbf{X}}, \hat{\mathbf{Y}}$), there are infinite unitary matrices $\mathbf{T}$ that satisfy: }
	
	\begin{equation}
	\hat{\mathbf{W}}^{\sharp{(i j k)}} \approx \hat{\mathbf{X}} \hat{\mathbf{Y}}^\top
	= (\hat{\mathbf{X}}\mathbf{T})(\mathbf{T}^\top \hat{\mathbf{Y}}^\top).
	\end{equation}
	
	{Among the unitary matrices, there is a rectification matrix $\hat{\mathbf{R}}$ that minimizes the distance between the auxilliary matrices ($\hat{\mathbf{X}}, \hat{\mathbf{Y}}$) and the ground truth (${\mathbf{X}}^{\star}, {\mathbf{Y}}^{\star}$):}\par
	
	\begin{equation}
	\underset{\hat{\mathbf{R}}}{\min}
	||{\mathbf{X}}\hat{\mathbf{R}}-{\mathbf{X}}^{\star}||^2_\mathbf{F}+
	||{\mathbf{Y}}\hat{\mathbf{R}}-{\mathbf{Y}}^{\star}||^2_\mathbf{F}.
	\end{equation}

	\begin{theorem}
		\label{Theorem_1}
		{On condition that each pixel of the observed image $\mathbf{W}$ only has i.i.d. Gaussian noise ($\mathbf{G}^{(i j k)} \stackrel{\mathrm{i.i.d.}}{\sim} \mathcal{N}(0, \sigma_0^{2})$ in Eq. (\ref{Eq_Wsharp_noise})) and the low-rank solver achieves the global minimum, the differences between the rectified auxiliary matrices estimations $\hat{\mathbf{X}}$ and $\hat{\mathbf{Y}}$ and the ground truth $\mathbf{X}^{\star}$ and $\mathbf{Y}^{\star}$ are:}
		\begin{equation}
		\label{Eq_ZxZyDefine}
		\begin{array}{l}{\hat{\mathbf{X}} \hat{\mathbf{R}}-\mathbf{X}^{\star}=\mathbf{Z}_{\mathbf{X}}+\Psi}_X \\ {\hat{\mathbf{Y}} \hat{\mathbf{R}}-\mathbf{Y}^{\star}=\mathbf{Z}_{\mathbf{Y}}+\Psi}_Y\end{array},
		\end{equation}
		\noindent where the unitary matrix $\hat{\mathbf{R}} \in \mathbb{R}^{r \times r}$ ($\hat{\mathbf{R}} \hat{\mathbf{R}}^{\top}=\mathbf{I}$) is the rectification matrix to rectify ($\hat{\mathbf{X}}$, $\hat{\mathbf{Y}}$) to ($\mathbf{X}^{\star}$, $\mathbf{Y}^{\star}$). { $\Psi_X \in \mathbb{R}^{K \times r}, \Psi_Y \in \mathbb{R}^{L \times r}$ are the negligible residuals.} The rows of the error matrix $\mathbf{Z}_X \in \mathbb{R}^{K \times r}$ (resp. $\mathbf{Z}_Y \in \mathbb{R}^{L \times r}$) are independent and obey:
		\begin{equation}
		\begin{array}{ll}{\mathbf{Z}_X^{\top} \mathbf{e}_v \stackrel{\text { i.i.d. }}{\sim} \mathcal{N}(0, \sigma_0^{2}({\Sigma}^{\star})^{-1}),} & {\text { for } 1 \leq j \leq K} \\ {\mathbf{Z}_Y^{\top} \mathbf{e}_v \stackrel{\text { i.i.d. }}{\sim} \mathcal{N}(0, \sigma_0^{2}({\Sigma}^{\star})^{-1}),} & {\text { for } 1 \leq j \leq K}\end{array},
		\end{equation}
		\noindent where $\mathbf{e}_v$ is the basis vector.
	\end{theorem}

	We provide the proof of \textbf{theorem} \ref{Theorem_1} in Section \ref{Section_appendix}. This proof is specifically designed for the full HSI patch denoising scenario and is clearer and more concise than the one in \cite{chen2019inference}. \par

	In \textbf{theorem} \ref{Theorem_1}, we make two assumptions, i.e., the Gaussian noise distribution and the global minimum obtained from the low-rank solver. The former ensures the correctness of the error propagation equations. The latter ensures that the residual noises $\Psi_X$ and $\Psi_Y$ are negligible compared to $\mathbf{Z}_{\mathbf{X}}$ and $\mathbf{Z}_{\mathbf{Y}}$, and Eq. (\ref{Eq_ZxZyDefine}) is approximated to:\par 
	\begin{equation}
	\label{Eq_ZxZyDefine_1}
	\begin{array}{l}{\hat{\mathbf{X}} \hat{\mathbf{R}}-\mathbf{X}^{\star}\approx\mathbf{Z}_{\mathbf{X}}} \\ {\hat{\mathbf{Y}} \hat{\mathbf{R}}-\mathbf{Y}^{\star}\approx\mathbf{Z}_{\mathbf{Y}}}\end{array}.
	\end{equation}

	Assuming that the first-order expansion is reasonably tight and omitting the high order expansion, the element-wise error between the estimation $\hat{\mathbf{W}}_{uv}^{\sharp{(i j k)}}$ and the ground truth ${\mathbf{W}}_{uv}^{\sharp\star{(i j k)}}$ at the position $uv$ is:

	\begin{equation}
	\label{Eq_error_S_Ssharp}
	\begin{aligned} 
	&\hat{\mathbf{W}}_{uv}^{\sharp{(i j k)}}-{\mathbf{W}}_{uv}^{\sharp\star{(i j k)}}\\
	&=[\mathbf{e}_{u}^{\top}\hat{\mathbf{X}}\hat{\mathbf{R}}(\hat{\mathbf{Y}} \hat{\mathbf{R}})^{\top}\mathbf{e}_{v}-\mathbf{e}_{u}^{\top}(\mathbf{X}^{\star} {\mathbf{Y}^{\star\top}}) \mathbf{e}_{v}]\\
	&=[\mathbf{e}_{u}^{\top}(\hat{\mathbf{X}}\hat{\mathbf{R}}-\mathbf{X}^\star+\mathbf{X}^\star)(\hat{\mathbf{Y}} \hat{\mathbf{R}}-\mathbf{Y}^\star+\mathbf{Y}^\star)^{\top}\mathbf{e}_{v}-\mathbf{e}_{u}^{\top}(\mathbf{X}^{\star} {\mathbf{Y}^{\star\top}}) \mathbf{e}_{v}]\\
	&\approx[\mathbf{e}_{u}^{\top}(\hat{\mathbf{X}} \hat{\mathbf{R}}-\mathbf{X}^{\star}) \mathbf{Y}^{\star \top} \mathbf{e}_{v}+\mathbf{e}_{u}^{\top} \mathbf{X}^{\star}(\hat{\mathbf{Y}} \hat{\mathbf{R}}-\mathbf{Y}^{\star})^\top \mathbf{e}_{v}] \\ 
	& = [\mathbf{e}_{u}^{\top} (\mathbf{Z}_X {\mathbf{Y}^{\star\top}}) \mathbf{e}_{v}+\mathbf{e}_{u}^{\top} (\mathbf{X}^{\star} \mathbf{Z}_Y^{\top}) \mathbf{e}_{v}],\\
	\end{aligned}
	\end{equation}
	
	\noindent where the bases $\mathbf{e}_u,\ \mathbf{e}_v$ localize the elements involved in calculating the element in position $(u,v)$. After some manipulation, we have the element-wise variance of the error as:
	\begin{equation}
	\label{Eq_U_V}
	\begin{aligned}
	&{\operatorname{var}}(\hat{\mathbf{W}}_{uv}^{\sharp{(i j k)}}-{\mathbf{W}}_{uv}^{\sharp\star{(i j k)}})\\
	& \stackrel{(\mathrm{i})}{\approx} [{\operatorname{var}}(\mathbf{e}_{u}^{\top} (\mathbf{Z}_X {\mathbf{Y}^{\star\top}}) \mathbf{e}_{v})+{\operatorname{var}}(\mathbf{e}_{u}^{\top} (\mathbf{X}^{\star} \mathbf{Z}_Y^{\top}) \mathbf{e}_{v})] &&\\
	&\stackrel{(\text { ii })}{=} 
	\sigma_0^{2}[{\mathbf{e}_{v}}^{\top} \mathbf{Y}^{\star}(\Sigma^{\star})^{-1} {\mathbf{Y}^{\star\top}} {\mathbf{e}_{v}}+
	\mathbf{e}_{u}^{\top} \mathbf{X}^{\star}(\Sigma^{\star})^{-1} \mathbf{X}^{\star \top} \mathbf{e}_{u}]\\ 
	&\stackrel{(\text { iii })}{=} \sigma_0^{2}(\|\mathbf{U}_{u,.}^{\star}\|_{2}^{2}+\|\mathbf{V}_{v,.}^{\star}\|_{2}^{2})
	= \sigma_0^{2}{v}_{uv}^{\star},
	\end{aligned}
	\end{equation}
	\noindent where (i) is derived from \textbf{theorem} \ref{Theorem_1} since $\mathbf{Z}_X$ and $\mathbf{Z}_Y$ are nearly independent. (ii) is obtained from Eq. (\ref{Eq_ZxZyDefine}). (iii) $\mathbf{U}_{u,.}^{\star}$ and $\mathbf{V}_{v,.}^{\star}$ are the $u$/$v$th row of $\mathbf{U}^{\star}$ and $\mathbf{V}^{\star}$. For conciseness, we define ${v}_{uv}^{\star}=(\|\mathbf{U}_{u,.}^{\star}\|_{2}^{2}+\|\mathbf{V}_{v,.}^{\star}\|_{2}^{2})$, and ${\operatorname{var}}(\cdot)$ is the variance.\par

	In practice, the rows of the ground truth $\mathbf{U}_{u,.}^{\star}$ and $\mathbf{V}_{v,.}^{\star}$ are impossible to obtain. Assuming that the solver achieves a solution close to the global minimum, we approximate them with the estimated version $\hat{\mathbf{U}}_{u,.}$ and $\hat{\mathbf{V}}_{v,.}$, which are the $u$/$v$th row of the estimations $\hat{\mathbf{U}}$ and $\hat{\mathbf{V}}$. Thus the element-wise variance can be approximated as:\par 
	
	\begin{equation}
	\begin{aligned}
	{\operatorname{var}}(\hat{\mathbf{W}}_{uv}^{\sharp{(i j k)}}-{\mathbf{W}}_{uv}^{\sharp\star{(i j k)}})
	{\approx} \sigma_0^{2}(\|\hat{\mathbf{U}}_{u,.}\|_{2}^{2}+\|\hat{\mathbf{V}}_{v,.}\|_{2}^{2}){=} \sigma_0^{2}\hat{v}_{uv},
	\end{aligned}
	\end{equation}
	\noindent where $\hat{v}_{uv}$ can be regarded as the estimated version of ${v}_{uv}^{\star}$. Due to the relationship of $\mathbf{W}^{\sharp\star(ijk)}_{uv}$ and $\hat{\mathbf{W}}_{uv}^{\sharp{(ijk)}}$, the corresponding variance of $\hat{\mathbf{W}}^{(ijk)}_{uv}$ is Eq. (\ref{Eq_uncertainty_quantification_exact}) and provides a closed-form variance describing the confidence of $\hat{\mathbf{W}}^{(ijk)}_{uv}$. The variance of $\hat{\mathbf{W}}^{(ijk)}_{uv}$ is:\par 
	
	\begin{equation}
	\label{Eq_uncertainty_quantification_exact}
	{\sigma_{\xi}^{(ijk)}}^2={\operatorname{var}}(\hat{\mathbf{W}}^{(ijk)}_{uv}-\mathbf{W}_{uv}^{\star{(ijk)}} )\approx g^{-1}(\sigma_0^{2}\hat{v}_{uv}),
	\end{equation}
	\noindent where $g^{-1}(\cdot)$ is the inverse permutation function of $g(\cdot)$. $\xi \subseteq\{1, \cdots, M\} \times\{1, \cdots, N\} \times\{1, \cdots, P\}$ is the corresponding 3D index of the 2D $uv$ at patch $(ijk)$.\par

	\subsection{Uncertainty estimation of the sliding-window}
	\label{subsection_slidingwindow}
	Another unique characteristic of the uncertainty quantification in the LRMA technique is the ``uncertainty mixing'' caused by the sliding-window strategy. The sliding window approach is seldom used in the 2D image domain, and recent work \cite{chen2019inference} did not address this issue. Thus we propose a method to solve the ``uncertainty mixing'' issue in HSI denoising due to the sliding-window strategy. Unlike averaging all patches to generate the final restored HSI (Eq. (\ref{fig:framework_uncertainty_RS})), the uncertainty of the product does not follow the typical error propagation, such as estimating the uncertainty of multiple observations \cite{casella2002statistical}. In the scenario of uncertainty estimation for i.i.d. observations, if the variance of each observation is $\sigma_o^2$ and we conduct $\mathrm{n}$ times observation, the variance of the averaged observation is $\frac{1}{n}\sigma_o^2$. This shows that the accuracy of the average observation is $n$ times more accurate than every single observation. However, in the LRMA process, we only make one observation. If we can process the entire HSI using one-time LRMA and the accuracy is $\sigma_o^2$, the sliding-window strategy in LRMA should not exceed $\sigma_o^2$ because no new information has been introduced. In other words, $\frac{1}{n}\sigma_o^2$ is not the correct uncertainty. The difference between multiple i.i.d. observations and the single processing scenarios is that the former has i.i.d. observations for all events, whereas the observations in the latter are correlated. In other words, the covariance between the overlapped patches in the LRMA is not null. For two related events $X$ and $Y$, the ratio $\eta$ is often used to describe the relationship:
	\begin{equation}
	\eta=\frac{\operatorname{cov}(X, Y)}{\sqrt{\operatorname{var}(X) \cdot \operatorname{var}(Y)}},
	\end{equation}
	\noindent where $\operatorname{cov}(\cdot)$ defines the covariance of the two events. If $\eta$ is 0, the events $X$ and $Y$ are i.i.d. If $\eta$ is 1, the events $X$ and $Y$ are linearly correlated, having the highest correlation.\par 
	
	We provide a toy example as the weighted average of events $X$ and $Y$ as $F=\mathrm{a} X+\mathrm{b} Y$. Assuming the variance of $X$ and $Y$ are $\sigma_X^2$ and $\sigma_Y^2$, the propagated uncertainty of $F$ is $\sigma_{F}^{2}$:

	\begin{equation}
	\sigma_{F}^{2}=\mathrm{a}^{2} \sigma_{X}^{2}+\mathrm{b}^{2} \sigma_{Y}^{2}+2 \eta_{XY} \mathrm{a} \mathrm{b}  \sigma_{X}\sigma_{Y},
	\label{Eq_classical_covariance}
	\end{equation}
	
	\noindent where $\eta_{XY}$ is the variance ratio defining the relation between events $X$ and $Y$. \par
	
	For convenience, we rewrite Eq. (\ref{Eq_image_stacking}) to the element-wise form:
	
	\begin{equation}
	\label{Eq_image_stacking_pixel}
	\begin{aligned}
	&\hat{\mathbf{W}}_{\xi}=\frac{1}{\phi(\xi)}\sum_{(i,j,k) \in \Phi(\xi)}\mathit{f}_p({\hat{\mathbf{W}}^{(ijk)})_{uv}}\\
	& \text{such that\ } (u,v) = {G}(\xi),
	\end{aligned}
	\end{equation}
	
	\noindent where $\Phi(\xi)$ defines the index of the patches covering the position $(u,v)$ and $\phi(\xi)$ is the number of those patches. Similar to the permutation function $g(\cdot)$, $G(\cdot)$ converts the 3D index of $\hat{\mathbf{W}}$ to the 2D premutated version $\hat{\mathbf{W}}^{\sharp}$. By denoting the element-wise variance of each $\mathit{f}_p({\hat{\mathbf{W}}^{(ijk)})_{uv}}$ is $\sigma^{(ijk)}_{uv}$, based on the uncertainty propagation theory, we obtain the variance that describes $\hat{\mathbf{W}}_{uv}$ as:
	
	\begin{equation}
	\label{Eq_sigma_slidingwindow}
	\begin{aligned} 
	\hat{\sigma}_{\xi}^2
	&=\frac{1}{\phi(\xi)^2}\sum_{(i,j,k) \in \phi(\xi)}{\sigma_{\xi}^{(ijk)}}^2+\\
	&\frac{1}{\phi(\xi)^2}\sum_{(i,j,k) \in \phi(\xi)}{\sum_{(i',j',k') \in \phi(\xi)}{2 \eta^{(ijk)(i'j'k')} {\sigma^{(ijk)}_{\xi}} {\sigma^{(i‘j’k‘)}_{\xi}}}},
	\end{aligned} 
	\end{equation}
	
	\noindent where each $(i,j,k)$ is different from $(i',j',k')$. $\eta^{(ijk)(i'j'k')}$ defines the relation between the element $\xi$ in patch $(ijk)$ and patch $(i'j'k')$. Readers may be aware that this is an extended multiple event version of the toy example in Eq. (\ref{Eq_classical_covariance}).\par
	
	We provide a practical solution to estimate the optimal ratio $\eta^{(ijk)(i'j'k')}$. {Similar to \textbf{theorem} \ref{Theorem_1}, we assume that the adopted low-rank solver (GoDec algorithm in this paper, but it could be any solver) achieves the global minimum, the information on the denoised element $\xi$ in the patch $\hat{\mathbf{W}}^{(ijk)}$ is determined by the sum of the information embedded in all the elements in the patch $\hat{\mathbf{W}}^{(ijk)}$.} The mutual information encoded in two overlapping patches is linearly related to the number of elements shared by the overlapping patches \cite{gray2011entropy}. Thus the mutual information of the element from two patches is linearly related to the number of shared elements in the two patches. Therefore, the optimal ratio between the two patches $\eta^{(ijk)(i'j'k')}$ is close to:
	
	\begin{equation}
	\label{Eq_mutual_info}
	\eta^{(ijk)(i'j'k')} {\to} \left[\frac{\mathit{N}((ijk),(i'j'k'))}{KL}\right]^{-},
	\end{equation}
	
	\noindent where $\mathit{N}((ijk),(i'j'k'))$ defines the number of shared elements in patch $(ijk)$ and patch $(i'j'k')$. $\left[\cdot \right]^{-}$ means slightly below $\cdot$. The ratio on the left is close to but smaller than that of the right term. The result is attributed to the assumption of a global minimum. {If the global minimum is guaranteed,  we have the information in each patch fully exploited; thus, the ratio is linearly related to the overlap.} Thus, the mutual information and number of elements are not strictly linearly related. Extreme case is the full overlap ($\eta^{(ijk)(i'j'k')}=1$) and no overlap ($\eta^{(ijk)(i'j'k')}=0$). Otherwise, if $\Psi_X$ and $\Psi_Y$ are not ignored, or the low-rank solver does not converge adequately, the variance based on the local minimum is not accurate and behaves more randomly; thus, the ratio is less than the expected right term. In the extreme case when the low-rank solver fails and converges randomly either due to bad performance of the algorithm or heavy random noise, the $\eta^{(ijk)(i'j'k')}$ should be 0. All other things being equal, a smaller ratio will lead to a smaller variance and slight over-confidence. However, a quantitative assessment of the approximation of Eq. (\ref{Eq_mutual_info}) is difficult because the performance of the low-rank solver remains difficult to quantify. \par
	
\begin{figure}[!h]
		\centering
		\centering
		\includegraphics[width=0.5\linewidth]{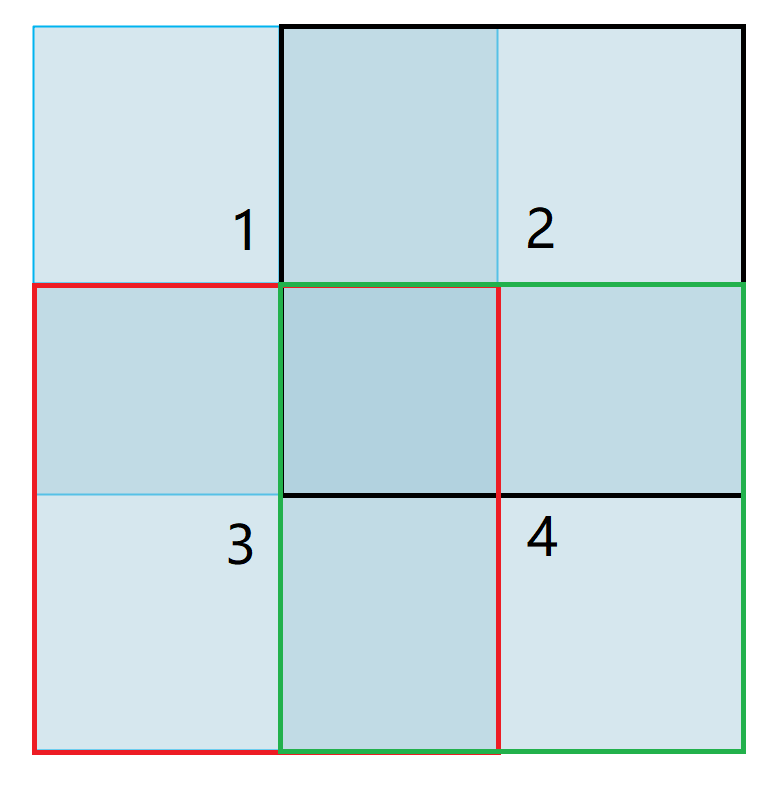}
		\caption{An example to explain the definition of mutual information defined in Eq. (\ref{Eq_mutual_info}). Each of the four 2D overlapping patches 1, 2, 3, and 4 has four pixels.}
		\label{fig:diagram_eta}
	\end{figure}

	Fig. \ref{fig:diagram_eta} presents an example of Eq. (\ref{Eq_mutual_info}). Each neighboring block pair has $50\%$ overlap. Based on the statistics of the number of overlapping elements, the $\eta$ of the pairs $(1,2)$, $(2,3)$, $(3,4)$ and $(4,1)$ is 0.5, whereas the $\eta$ of pairs $(1,3)$ and $(2,4)$ is 0.25. \par 
	
	
	\subsection{The impact of random noise}
	\label{Sec_impact_randomnoise}
	The result of Eq. (\ref{Eq_uncertainty_quantification_exact}) is based on the assumption that the noises $\Psi_X$ and $\Psi_Y$ are significantly smaller than the variables $\mathbf{Z}_{\mathbf{X}}$ and $\mathbf{Z}_{\mathbf{Y}}$ with a normal distribution. If $\Psi_X$ and $\Psi_Y$ are not ignored, the conclusion does not hold. Therefore, it remains challenging to quantify the impact of random noise versus Gaussian noise. However, as \cite{wright2009robust} pointed out, the RPCA-like formulation (Eq. (\ref{Eq_image_recovery}).) is robust against random noise since the low-rank solver will adjust the use of the norm-1 or norm-2 to regularize the state matrix. Correspondingly, elements not corrupted by random noise are unaffected, whereas the impulse noise can be identified. Therefore, regarding the impact of random noise, the robustness of our approach is closely related to the choice of the LRMA algorithm. It should be noted that most existing HSI denoising studies have focused on modeling Gaussian noise \cite{zhuang2018fast, bioucas2008hyperspectral}. A few LRMA-based methods have addressed non-Gaussian noise \cite{he2015total}. These approaches utilize a total-variance related term to filter out non-Gaussian noise in the HSI to maintain good boundaries after filtering. Similarly, the proposed uncertainty estimation method can also be integrated with the method in \cite{he2015total} since the total-variance term is only a constraint.\par

	\begin{figure*}[!h]
		\centering
		\subfloat[]{
			\begin{minipage}[]{0.25\textwidth}
				\centering
				\includegraphics[width=1\linewidth]{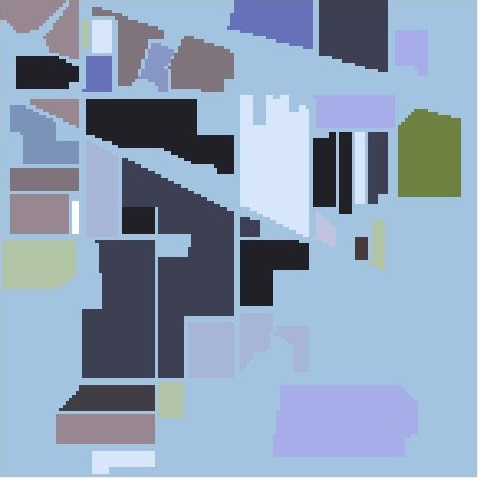}
			\end{minipage}
		}
		\subfloat[]{
			\begin{minipage}[]{0.45\textwidth}
				\centering
				\includegraphics[width=1\linewidth]{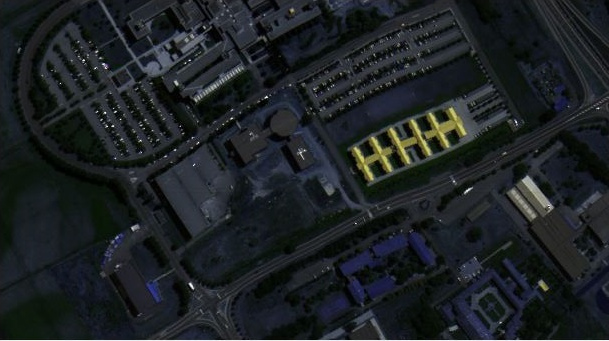}
			\end{minipage}
		}
		\subfloat[]{
			\begin{minipage}[]{0.25\textwidth}
				\centering
				\includegraphics[width=1\linewidth]{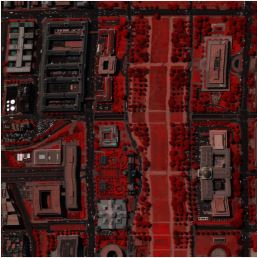}
			\end{minipage}
		}
		\caption{(a) False-color image of the Indian Pines dataset using bands 10, 37, and 70. This is the manually classified ground truth of the Indian Pines dataset. (b) False-color image of the University of Pavia dataset composed of the bands 20, 42, and 72. (c) False-color image of the Washington D.C dataset composed of bands 60, 27, and 17.}
		\label{fig:dataset_used}
	\end{figure*}

	\begin{figure*}[!h]
		\centering
		\subfloat[Indian dataset. $\sigma_0=0.050$]{
			\begin{minipage}[]{0.23\textwidth}
				\centering
				\includegraphics[width=1\linewidth]{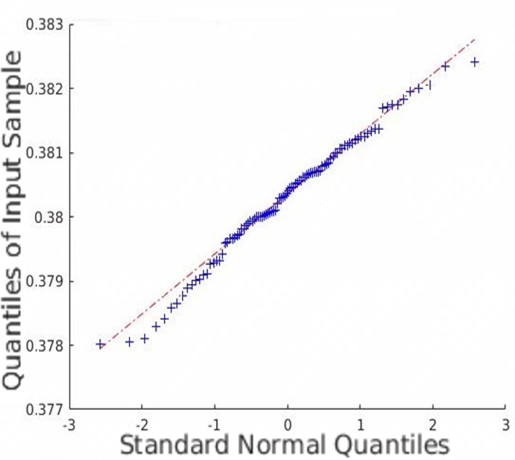}
			\end{minipage}
		}
		\subfloat[Indian dataset. $\sigma_0=0.075$]{
			\begin{minipage}[]{0.23\textwidth}
				\centering
				\includegraphics[width=1\linewidth]{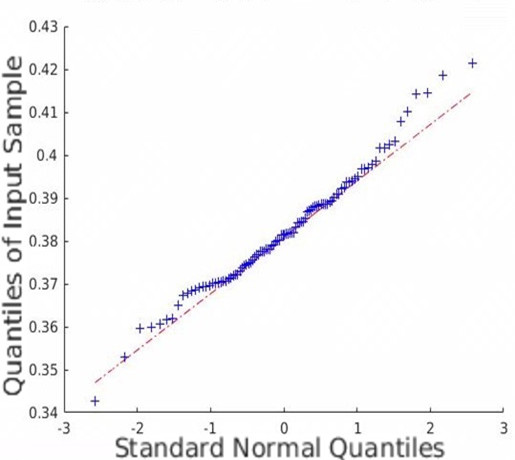}
			\end{minipage}
		}
		\subfloat[Indian dataset. $\sigma_0=0.100$]{
			\begin{minipage}[]{0.23\textwidth}
				\centering
				\includegraphics[width=1\linewidth]{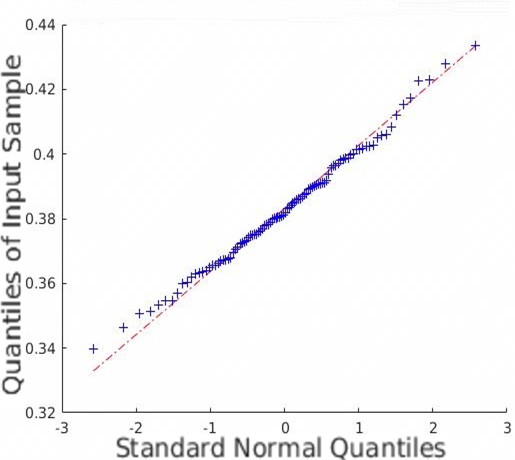}
			\end{minipage}
		}
		\subfloat[Indian dataset. $\sigma_0=0.125$]{
			\begin{minipage}[]{0.23\textwidth}
				\centering
				\includegraphics[width=1\linewidth]{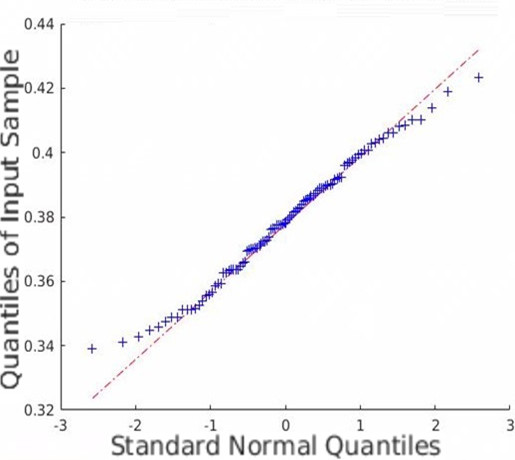}
			\end{minipage}
		}\\
		\subfloat[University of Pavia dataset. $\sigma_0=0.050$]{
			\begin{minipage}[]{0.23\textwidth}
				\centering
				\includegraphics[width=1\linewidth]{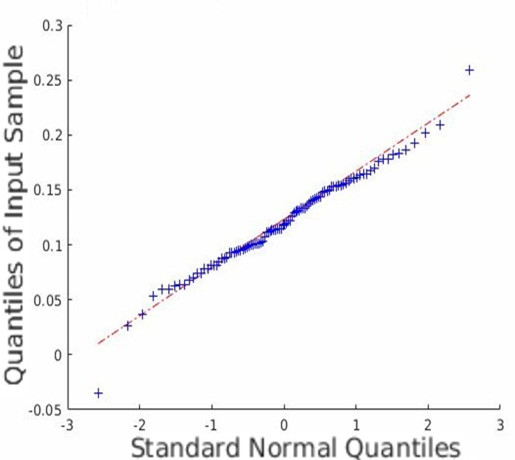}
			\end{minipage}
		}
		\subfloat[University of Pavia dataset. $\sigma_0=0.075$]{
			\begin{minipage}[]{0.23\textwidth}
				\centering
				\includegraphics[width=1\linewidth]{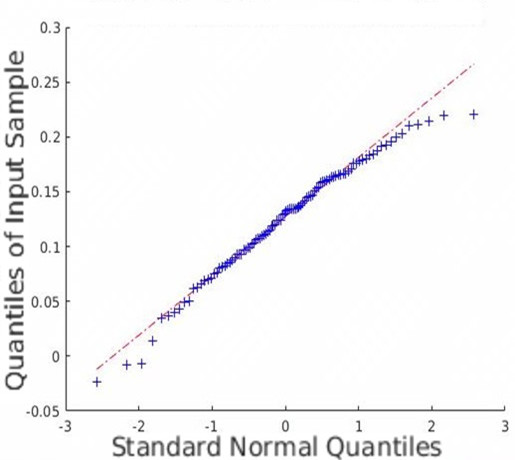}
			\end{minipage}
		}
		\subfloat[University of Pavia dataset. $\sigma_0=0.100$]{
			\begin{minipage}[]{0.23\textwidth}
				\centering
				\includegraphics[width=1\linewidth]{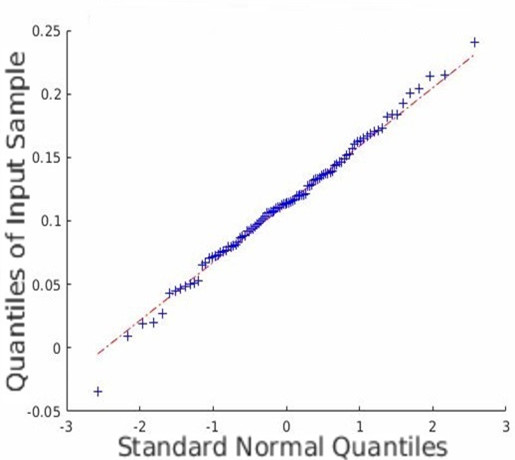}
			\end{minipage}
		}\/
		\subfloat[University of Pavia dataset. $\sigma_0=0.125$]{
			\begin{minipage}[]{0.23\textwidth}
				\centering
				\includegraphics[width=1\linewidth]{QQ_Indian_125.JPG}
			\end{minipage}
		}\\
		\subfloat[Washington,D.C. $\sigma_0=0.050$]{
			\begin{minipage}[]{0.23\textwidth}
				\centering
				\includegraphics[width=1\linewidth]{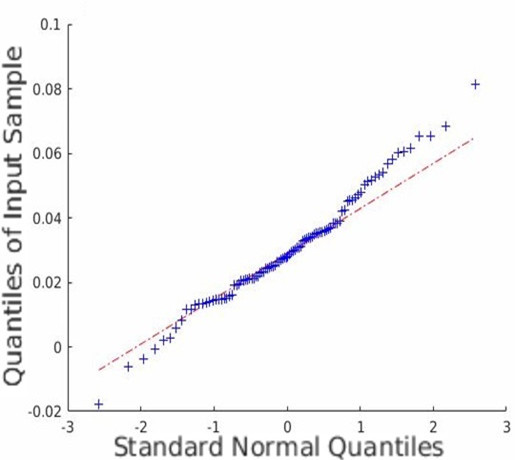}
			\end{minipage}
		}
		\subfloat[Washington,D.C. $\sigma_0=0.075$]{
			\begin{minipage}[]{0.23\textwidth}
				\centering
				\includegraphics[width=1\linewidth]{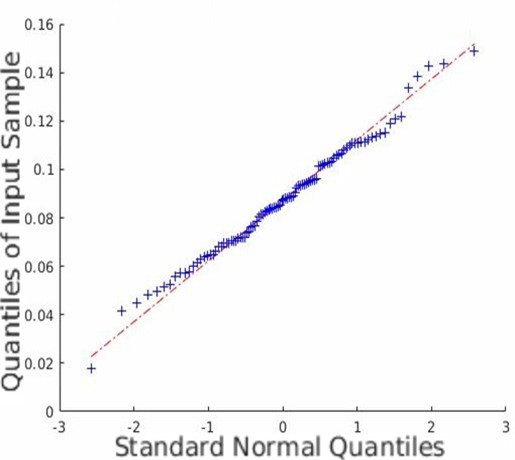}
			\end{minipage}
		}
		\subfloat[Washington,D.C. $\sigma_0=0.100$]{
			\begin{minipage}[]{0.23\textwidth}
				\centering
				\includegraphics[width=1\linewidth]{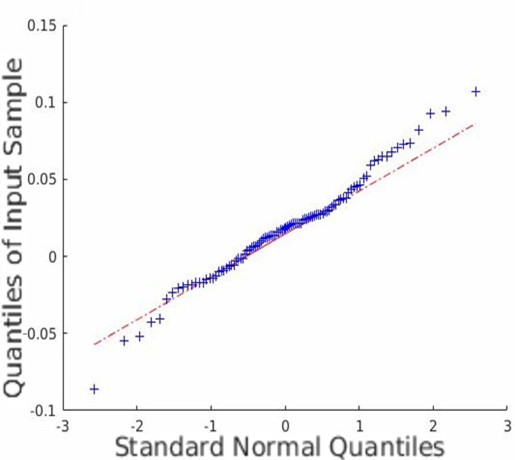}
			\end{minipage}
		}\/
		\subfloat[Washington,D.C. $\sigma_0=0.125$]{
			\begin{minipage}[]{0.23\textwidth}
				\centering
				\includegraphics[width=1\linewidth]{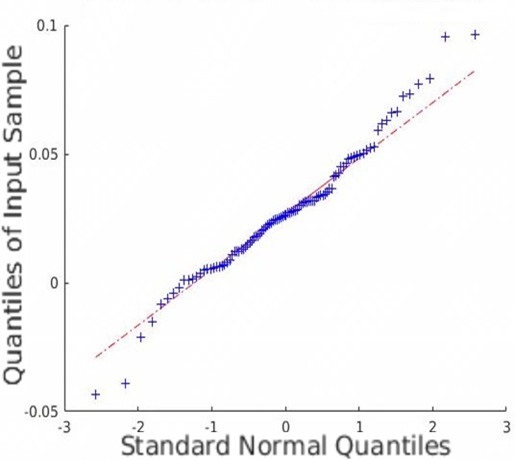}
			\end{minipage}
		}
		\caption{The Q-Q plots of a random element position in 100 independent trials to assess the noise in the three datasets. }
		\label{fig:3datasets}
	\end{figure*}

	\section{Results and discussion}
	\label{Section_results}
	
	Our work aims at providing an uncertainty quantification approach and does not consider convergence analysis, hyperparameter definition, denoised image visualization, or comparison of HSI denoising methods. For clarifications on the methods described in Section \ref{Section_method}, we refer the reader to \cite{zhang2013hyperspectral,he2015total,chen2017denoising,zhuang2018fast,fan2018spatial,chen2019hyperspectral}. This work is NOT a new LRMA formulation. We only demonstrate that the proposed closed-form uncertainty estimation approach (Section \ref{Section_uncertianty_quantify}) quantifies the denoised HSI accurately, efficiently, and robustly. Since previous works have not provided any uncertainty quantification methods, we only show that our proposed approach can statistically describe the uncertainty based on the Monte Carlo tests.\par 
	
	To the best of our knowledge, a Monte Carlo test is the only appropriate method to validate the efficiency of the proposed method. The complexity of the LRMA formulation hinders the closed-form retrieval of the posterior distribution of the denoised image. This problem motivated us to provide the closed-form estimation of the variance. Therefore, we could not design an experiment providing the ground truth information on both the noisy image and the denoised image. Moreover, regarding the Monte Carlo tests, we should emphasize that the noise of the original image follows identical Gaussian distribution while noise of the denoised image is non-identical. The noise of any pixel in original observations follows $\mathbf{G}^{(i j k)} \stackrel{\mathrm{i.i.d.}}{\sim} \mathcal{N}(0, \sigma_0^{2})$. However, as Eq. (\ref{Eq_uncertainty_quantification_exact}) indicates, the noise of the denoised image depends on the ratio $\hat{v}_{uv}$. Thus, the pixels of the denoised image do not maintain the same noise level. The simulated identical noise can only be enforced on the original image. The Monte Carlo test is the optimal method to obtain the posterior distribution of the denoised image. Theoretically, the Monte Carlo test can generate the ground truth variance when the number of trials is large enough.\par
	
	We validate the proposed method on the Indian Pines dataset (simulation dataset) \footnote{https://engineering.purdue.edu/~biehl/MultiSpec/hyperspectral.html}, the University of Pavia dataset (real dataset) \footnote{http://www.ehu.eus/ccwintco/index.php?title=Hyperspectral \\ \_Remote\_Sensing\_Scenes}, and the Washington, D.C. mall dataset \footnote{https://engineering.purdue.edu/~biehl/MultiSpec/hyperspectral.html} (real dataset). The Indian Pines dataset is a subset of the data acquired by the AVIRIS sensor. It consists of $145\times145$ pixels and 224 spectral reflectance bands in the wavelength range of $0.4–2.5 10^{-6}$ meters and covers the Indian Pines test site in northwestern Indiana, USA. The Indian Pines scene consists of two-thirds agricultural land and one-third forest or other natural perennial vegetation. We use the ground truth of the Indian Pines dataset and simulate arbitrary Gaussian noise. The University of Pavia dataset was acquired by the ROSIS sensor and covers Pavia in northern Italy. It consists of $610\times610$ pixels with 103 spectral bands; however, some of the samples in the images contained no information and were discarded before the analysis. The geometric resolution is 1.3 meters. We select a subset of $610\times610\times103$ in the experiments. In the Washington, D.C. mall dataset, the image contains 191 spectral bands with a size of $1208 \times307 $. We clip the image to a size of $256\times 256\times 191$ in our experiment. Fig. \ref{fig:dataset_used} shows the false-color images of the three datasets.\par

	We strictly follow the parameter and noise setting of the selected LRMA method to provide a convincing validation \cite{zhang2013hyperspectral}. The size of the sliding window is $20 \times 20 \times P$, where $P$ is the number of bands in the dataset. The step size is 4, and the rank $\mathrm{r}$ is 7. In the image restoration process, all images are normalized to the range $[0\ 1]$. The zero-mean Gaussian noise is uniformly applied to all pixels in the datasets with different levels of noise. Since the two real datasets do not have ground truth, we use the output of \cite{zhang2013hyperspectral} as the ground truth. The impact of impulse noise is further tested on the datasets. It should be noted that this research is based on the known raw HSI noise, which can be retrieved using previously described methods \cite{aiazzi2006noise,acito2011signal}.\par

	\begin{table}[]
		\caption{The coverage rates of ${\operatorname{var}}(\hat{\mathbf{W}}_{\xi}^{\{l\}}-\overline{\hat{\mathbf{W}}}_{\xi} )$ for different $\sigma_0$ for 100 Monte Carlo trials. ``Std'' denotes the standard deviation. The proposed closed-form uncertainty estimation approach describes the probability distribution well.} 
		\centering
		\begin{tabular}{c|c|c|c|c|c|c}
			\hline
			& \multicolumn{2}{c|}{Indian} & \multicolumn{2}{c|}{University of Pavia} & \multicolumn{2}{c}{Washington, D.C.} \\ \hline
			$\sigma_0$ & Mean         & Std          & Mean                & Std                & Mean              & Std              \\
			0.025      & 0.9514       & 0.0483       &   0.9573    &       0.0515       &         0.9411    &      0.0583       \\
			0.050      & 0.9579       & 0.0371       & 0.9476              & 0.0292             & 0.9368            & 0.0348           \\
			0.075      & 0.9576       & 0.0383       & 0.9556              & 0.0263             & 0.9501            & 0.0311           \\
			0.100      & 0.9643       & 0.0319       & 0.9652              & 0.0233             & 0.9626            & 0.0266           \\
			0.125      & 0.9704       & 0.0273       & 0.9735              & 0.0205             & 0.9712            & 0.0297           \\ \hline
		\end{tabular}
		\label{table:comparison_of_coverage_ratio}
	\end{table}

	\begin{figure}[!h]
		\centering
		\centering
		\includegraphics[width=1\linewidth]{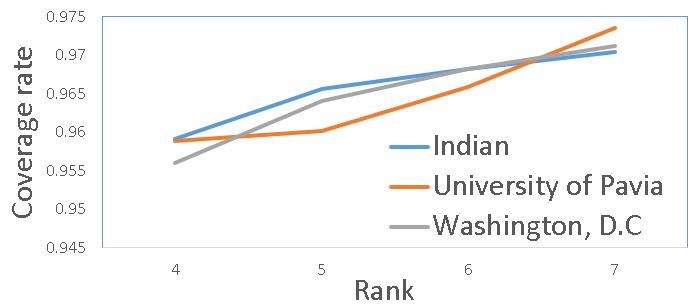}
		\caption{The relationship between the average coverage rate and the rank at the noise level of $\sigma_0=0.125$. As compliment to Table \ref{table:comparison_of_coverage_ratio} with fixed scale of noise, this figure provides a comparison of different level of noises.}
		\label{fig:cov_rank}
	\end{figure}
	
	\subsection{The validation of the uncertainty description}
	The accuracy of the proposed closed-form uncertainty quantification method is extensively validated with numerical Monte Carlo tests. We apply different levels of Gaussian noises to the HSI. The numerical experiments provide sufficient samples to test if the statistical distributions of the samples are described accurately with the output confidence interval. Specifically, $\mathrm{T}=100$ Monte Carlo tests are performed on the input of the same noise $\sigma_0^2$. We should emphasize that the 100 trials are chosen due to limited computational resources. Due to the large volume of the HSIs, 100 Monte Carlo trials are sufficient to obtain a statistic assessment of the efficiency of the proposed method. The code is provided for readers to conduct more tests. For each element of the estimated HSI in trial $l$ at position $\xi$ as $\hat{\mathbf{W}}_{\xi}^{\{l\}}$, we define its deviation from the corresponding $\mathrm{T}$ times mean value as $\overline{\hat{\mathbf{W}}}_{\xi}$. The deviation from the mean is defined as:
	
	\begin{equation}
	\label{Eq_deviatioin}
	\mathrm{e}^{\{l\}}_{\xi}=||\hat{\mathbf{W}}_{\xi}^{\{l\}}-\overline{\hat{\mathbf{W}}}_{\xi}||_1,\  l\in [1,\cdots,\mathrm{T}].
	\end{equation} 
	
	If the closed-form solution describes the statistical distribution of the estimation well, $95\%$ of $\hat{\mathbf{W}}_{\xi}^{\{l\}}$ should fall within the range of $\left[\overline{\hat{\mathbf{W}}}_{\xi} - 1.96\hat{\sigma}_{\xi},\ \overline{\hat{\mathbf{W}}}_{\xi} +1.96\hat{\sigma}_{\xi}\right]$. Following Eq. (\ref{Eq_deviatioin}), we define the ratio of $\mathrm{e}^{\{l\}}_{\xi}$ falling within the $1.96\hat{\sigma}_{\xi}$ bound as the element-wise \textbf{coverage rate}. The average coverage rate of all elements is the \textbf{average coverage rate}. Table \ref{table:comparison_of_coverage_ratio} presents the average and standard deviation of the coverage rate. The result indicates that the uncertainty approach robustly and accurately describes the element-wise denoised HSI when the level of noise is low ($\sigma_0$ less than $0.100$) \footnote{\url{https://JingweiSong@bitbucket.org/JingweiSong/uncertainty-estimation-for-hyper-spectral-image-denoising.git}.}. In the case $\sigma_0 \geq 0.100$, we notice an under-confident of the prediction. We hypothesize that the accuracy is related to the level of noise and the choice of the optimal rank. When the noise level increases, the fixed rank LRMA does not accurately filter the noise while retaining useful information. A lower rank is needed to ensure that the residual follows the assumption of the normal distribution of the noise. As depicted in Table \ref{table:comparison_of_coverage_ratio}, we strictly follow \cite{zhang2013hyperspectral} by setting $\mathrm{r}=7$. To determine the reason, we conduct an experiment by choosing different $\mathrm{r}$. The results are shown in Fig. (\ref{fig:cov_rank}), indicating that the performance  (uncertainty quantification) of our approach improves by decreasing the arbitrary rank. This implies that the Monte Carlo experiment provides a method to validate the correct choice of the rank. Further studies are needed to determine the rank at high noise level.\par


	\subsection{The validation of the normal distribution of the restored HSI}
	In addition to the \textbf{average coverage rate}, we also adopt the Monte Carlo experiments to prove that the restored HSI $\hat{\mathbf{W}}^{\{l\}}$ has an innate Gaussian distribution. Quantile-Quantile (Q-Q) plots are created, and Shapiro-Wilk tests are conducted to validate the proposed approach. Fig. \ref{fig:3datasets} shows the Q-Q plots of all the datasets. Note that we randomly plot one element in each dataset as an example, but all other elements are similar since the observation is isotropic. Fig. \ref{fig:3datasets} indicates that the Monte Carlo estimation of all the trials $\hat{\mathbf{W}}^{\{l\}}$ is close to a line; thus, the sample follows the Gaussian distribution. \par

	In the Shapiro-Wilk tests, it is not possible to present all the elements. Thus, we randomly choose several elements from the three datasets with different levels of Gaussian noise. The results are presented in Table \ref{table:wilk_shapiro_test}. The $p$-values indicate that all samples follow the normal distribution ($> 0.05$). Since all pixels are acquired under the same conditions, and their statistics should be consistent (either obey or reject the normal distribution assumption), we can conclude that the restored HSI still follows the Gaussian distribution.\par

	\begin{table}[]
		\caption{The probability values ($p$-value) of the Shapiro-Wilk tests of the Indian Pines, University of Pavia, and Washington, D.C. mall datasets. We cite the corresponding element positions shown in Fig. (\ref{fig:3datasets}) as an example through all the tests. The $p$-value $> 0.05$ indicates a significant possibility that the element is normally distributed.} 
		\centering
		\begin{tabular}{l|l|c|c}
			\hline
			Dataset & Indian & University of Pavia & Washington, D.C. \\ \hline
			0.050   & 0.4239 & 0.1531              & 0.2046                        \\
			0.075   & 0.1103 & 0.1924              & 0.0585                        \\
			0.100   & 0.8498 & 0.7537              & 0.6977                        \\
			0.125   & 0.4346 & 0.1531              & 0.5331                        \\ \hline
		\end{tabular}
		\label{table:wilk_shapiro_test}
	\end{table}
	
	\subsection{Time consumption comparison}
	In addition to the accuracy tests, we determine the time consumption. We find that the closed-form uncertainty estimation method has similar time consumption as the LRMA and far less time consumption than the Monte Carlo numerical tests. We use the LRMA code used in \cite{zhang2013hyperspectral} and add the uncertainty estimation module. The proposed algorithm, the LRMA, and the Monte Carlo tests are all implemented in Matlab 2019a. In the Monte Carlo tests, we utilize the Parallel Computing Toolbox of Matlab to speed up the steps. The hardware set is $3.1 GHz$ Intel Xeon E5-2687wv3 and 64 GB DDR4-Ram. In the Monte Carlo experiments, all 20 CPU threads are utilized. Note that in the Monte Carlo tests, each element is covered by several sliding windows. Since the memory is not sufficient to store all sliding windows, we calculate the result of the patch, store it on the hard disk, and reload it to calculate each element in the final step.\par
	
	\begin{table}[]
		\caption{The time consumption of the Monte Carlo uncertainty estimation, the LRMA method \cite{zhang2013hyperspectral}, and the proposed method. Note that 100 trials are conducted in the Monte Carlo tests. The time is recorded in seconds.} 
		\centering
		\begin{tabular}{c|c|c|c}
			\hline
			Dataset             & Monte Carlo & LRMA & LRMA+Uncertainty \\ \hline
			Indian              & 7862        & 653  & 864                     \\
			University of Pavia & 8393        & 815  & 930                     \\
			Washington, D.C.     &   12437   & 1023 &   1590                   \\ \hline
		\end{tabular}
		\label{table:time_consumption}
	\end{table}
	
	Table \ref{table:time_consumption} shows the time consumption of the Monte Carlo uncertainty estimation, the LRMA method \cite{zhang2013hyperspectral}, and the proposed method integrated into \cite{zhang2013hyperspectral}. The 100 Monte Carlo trials are significantly speeded up because of the running in parallel using 20 CPU threads. The results indicate that the proposed closed-form estimation does not add a significant amount of time to the LRMA method, whereas the Monte Carlo test requires substantially more time. As indicated by Eq. (\ref{Eq_U_V}), the proposed approach requires no more than one band-wise SVD decomposition of the given HSI in addition to the original LRMA solver.

	\subsection{The impact of random noise}
	
	\begin{table*}[]
		\caption{The coverage rate and standard deviation of different ratios of impulse noise. We add $\mathrm{m}\%, \mathrm{m}\in [5,10,20,30]$ ratios of impulse noise to the observations, and test the coverage rates of different scenarios in 100 Monte Carlo trials. } 
		\centering
		\begin{tabular}{c|c|c|c|c|c|c|c|c|c}
			\hline
			&            & \multicolumn{2}{c|}{5\%} & \multicolumn{2}{c|}{10\%} & \multicolumn{2}{c|}{20\%} & \multicolumn{2}{c}{30\%} \\ \hline
			Dataset                                                                          & $\sigma_0$ & Mean        & Std        & Mean        & Std         & Mean        & Std         & Mean        & Std         \\ \hline
			\multirow{4}{*}{Indian}                                                          & 0.050      & 0.9006      & 0.0619     & 0.8429      & 0.0770      & 0.7309      & 0.0924      & 0.6278      & 0.0952      \\
			& 0.075      &   0.9348   &    0.0476   & 0.9063      & 0.0578      & 0.8395      & 0.0745      &     0.7621  &   0.0860   \\
			& 0.100      &  0.9516     &   0.0382   & 0.9340      &  0.0456     & 0.8888        &  0.0606   &  0.8293   &    0.0747  \\
			& 0.125      &  0.9624  &     0.0318   &       0.9495 &    0.0379   &   0.9148    &  0.0516     &   0.8657    & 0.0663    \\ \hline
			\multirow{4}{*}{\begin{tabular}[c]{@{}c@{}}University\\ of\\ Pavia\end{tabular}} & 0.050      & 0.9008      & 0.0438     & 0.8472      & 0.0558      & 0.7466      & 0.0711      & 0.6578      & 0.0783      \\
			& 0.075      & 0.9434      & 0.0343     & 0.9189      & 0.0432      & 0.8640      & 0.0591      & 0.8020      & 0.0715      \\
			& 0.100      & 0.9649      & 0.0275     & 0.9519      & 0.0340      & 0.9195      & 0.0477      & 0.8778      & 0.0607      \\
			& 0.125      & 0.9778      & 0.0220     & 0.9700      & 0.0270      & 0.9496      & 0.0382      & 0.9201      & 0.0503      \\ \hline
			\multirow{4}{*}{\begin{tabular}[c]{@{}c@{}}Washington, \\ D.C\end{tabular}}      & 0.050      &   0.9010 &    0.0549  &   0.8536   &    0.0690  &    0.7649   &    0.0860  &  0.6840    &  0.0933    \\
			& 0.075      & 0.9456  &   0.0417  &   0.9250  &  0.0514     &  0.8781   &    0.0689  &   0.8248  &    0.0835  \\
			& 0.100      &   0.9674  &  0.0319  &   0.9566  &  0.0388      &  0.9301     &   0.0534   &   0.8954  &   0.0688   \\
			& 0.125      &   0.9798    &   0.0245  &   0.9735   &    0.0297    &   0.9566   &   0.0417   &   0.9330     &   0.0560    \\ \hline
		\end{tabular}
		\label{table_robustness_analysis}
	\end{table*}
	
	In addition to the coverage rate tests, we assess if the proposed approach is robust to the random noise from the sensor. As Eq. (\ref{Eq_image_recovery0}) shows, our work ignores the random noise $\mathbf{N}$ and only assumes that the residuals follow the Gaussian noise $\mathbf{G}$. As described in Section \ref{Sec_impact_randomnoise}, we expect that the proposed method is robust to random noise such as impulse noise, dead lines, and stripes \cite{vane1993airborne,green1998imaging}. We conduct the tests by deliberately corrupting the original HSI with random noise and test the performance of the proposed uncertainty quantification method. We simulate different proportions of impulse noise on all datasets and determine the coverage rate of the Monte Carlo experiments.\par
	
	Table \ref{table_robustness_analysis} shows the results of the Monte Carlo experiments for different coverage rates. The results indicate a relationship between the robustness of the proposed closed-form method and the coverage rate. When the noise $\sigma_0$ is greater than $0.1$, the uncertainty quantification approach is less sensitive to pulse noises. When the noise $\sigma_0$ is less than $0.075$, the proposed method is susceptible to random noise. Theoretically, the proposed approach is only effective for Gaussian noise, which is the low-rank prior to Eq. (\ref{Eq_image_recovery0}). Therefore, less Gaussian noise means less reliable of the assumption that $\mathbf{N}$ is less than $\mathbf{G}$ in Eq. (\ref{Eq_observation_noise}). Thus, the results from the closed-form approach are less reliable. When the Gaussian noise level of the observation is low, the proposed approach tends to be over-confident of the result, and the result is lower than the coverage rate with a higher $\sigma_0$. A higher $\sigma_0$ means that the distribution is closer to the impulse noise. Furthermore, since the solvers (either GoDec or ALM solver) are all based on SVD decomposition, our approach has the same computational complexity as the original LRMA methods.\par

	\section{Conclusion}
	\label{Section_Con}
	This is the first study that addresses the uncertainty quantification of LRMA-based HSI denoising techniques. We provide a time-efficient closed-form uncertainty propagation model to quantify the element-wise uncertainty of the denoised HSI without requiring any ground truth. We also propose a weighted average uncertainty formulation to tackle the ``uncertainty mixing'' problem caused by the sliding-window strategy in HSI denoising. Moreover, the proposed uncertainty estimation approach is independent of the choice of the LRMA solver. It is based on the global minimum assumption of LRMA; thus, a more accurate algorithm leads to a better performance of the uncertainty estimation. Extensive numerical experiments indicate that the proposed uncertainty quantification method is accurate and robust and describes the statistical behavior of the estimates obtained from the LRMA approach. The time consumption of the closed-form formulation is low; therefore, the method can be deployed in any state-of-the-art LRMA algorithm. The proposed uncertainty approach can also be used in applications that require multi-source or multi-scale data fusion, data assimilation, and remote sensing uncertainty estimation. \par
	
	Future works may also focus on tackling the observable random noise, such as impulse noises, deadlines, and stripes. The percentage of the observable random noise can be utilized to adjust the proposed closed-form uncertainty method. \par

	\appendix
	
	\section{Appendix}
	\label{Section_appendix}
	
	\textbf{Proof of Theorem 1}. If the selected LRMA solver provides the optimal solution to the objective function Eq. (\ref{Eq_image_recovery}), the optimizer $\hat{\mathbf{X}}$, $\hat{\mathbf{Y}}$ should satisfy the local minima of Eq. (\ref{Eq_image_recovery}).  $\mathit{F}(\hat{\mathbf{X}},\hat{\mathbf{Y}})$ are denoted as:
	\begin{equation}
	\begin{aligned}
	\mathit{F}(\hat{\mathbf{X}},\hat{\mathbf{Y}})
	&=\frac{1}{2}\|\mathbf{W}^{(ijk)}-g^{-1}(\hat{\mathbf{X}}\hat{\mathbf{Y}}^\top)\|_\mathrm{F}^{2}\\
	&\stackrel{(\mathrm{i})}=\frac{1}{2}\|\mathbf{W}^{\sharp{(ijk)}}-\hat{\mathbf{X}}\hat{\mathbf{Y}}^\top\|_\mathrm{F}^{2},
	\end{aligned}
	\end{equation}
	\noindent where $(\mathrm{i})$ is because $g^{-1}(\cdot)$ is the permutation function and $g(\mathbf{W}^{(ijk)})=\mathbf{W}^{\sharp(ijk)}$. Assuming the solver yields  $F(\hat{\mathbf{X}},\hat{\mathbf{Y}})$ close to the global minimum $F(\mathbf{X}^{\star},\mathbf{Y}^{\star})$. The first-order expansion of $F(\hat{\mathbf{X}},\hat{\mathbf{Y}})$ is close to the zero matrix defined as $\mathcal{O}$. By solving Eq. (\ref{Eq_observation_noise}) and ignoring random noise, we define the patch function:
	
	\begin{equation}
	\begin{aligned}
	&\mathbf{W}^{(ijk)}=\mathbf{W}^{\star{(ijk)}}+\mathbf{G}^{(ijk)}\\
	&\mathbf{W}^{\sharp(ijk)}=\mathbf{W}^{\sharp\star{(ijk)}}+\mathbf{G}^{{\sharp(ijk)}},
	\end{aligned}
	\end{equation}
	
	\noindent where $\mathbf{G}^{{\sharp(ijk)}} = g(\mathbf{G}^{(ijk)})$. The optimizer $F(\hat{\mathbf{X}},\hat{\mathbf{Y}})$ achieves local minima which is close to the global minima; thus, we obtain the partial derivatives equal to the zero matrix:\par

	\begin{equation}
	\label{Eq_provePartialZero1}
	\begin{aligned}
	\frac{\partial{F(\hat{\mathbf{X}},\hat{\mathbf{Y}})}}{\partial{\hat{\mathbf{X}}}}&\stackrel{(\mathrm{i})}=(\mathbf{W}^{\sharp(ijk)}-\hat{\mathbf{X}}\hat{\mathbf{Y}}^\top)\hat{\mathbf{Y}}\\
	&=(\mathbf{G}^{\sharp(ijk)}+\mathbf{W}^{\star(ijk)}-\hat{\mathbf{X}}\hat{\mathbf{Y}}^\top)\hat{\mathbf{Y}}\\
	&=(\mathbf{G}^{\sharp(ijk)}+\mathbf{X}^\star {\mathbf{Y}^{\star\top}}-\hat{\mathbf{X}}\hat{\mathbf{Y}}^\top)\hat{\mathbf{Y}}\\
	&\stackrel{(\mathrm{ii})}\approx(\mathbf{G}^{\sharp(ijk)}+\mathbf{X}^\star\hat{\mathbf{R}}^\top \hat{\mathbf{Y}}^{\top}-\hat{\mathbf{X}}\hat{\mathbf{Y}}^\top)\hat{\mathbf{Y}}\approx\mathcal{O},\\
	\end{aligned}
	\end{equation}

	\begin{equation}
	\label{Eq_provePartialZero2}
	\begin{aligned}
	\frac{\partial{F(\hat{\mathbf{X}},\hat{\mathbf{Y}})}}{\partial{\hat{\mathbf{Y}}}}&\stackrel{(\mathrm{i})}=\hat{\mathbf{X}}^\top(\mathbf{W}^{\sharp(ijk)}-\hat{\mathbf{X}}\hat{\mathbf{Y}}^\top)\\
	&=\hat{\mathbf{X}}^\top(\mathbf{G}^{\sharp(ijk)}+\mathbf{W}^{\star(ijk)}-\hat{\mathbf{X}}\hat{\mathbf{Y}}^\top)\\
	&=\hat{\mathbf{X}}^\top(\mathbf{G}^{\sharp(ijk)}+\mathbf{X}^\star {\mathbf{Y}^{\star\top}}-\hat{\mathbf{X}}\hat{\mathbf{Y}}^\top)\\
	&\stackrel{(\mathrm{ii})}\approx\hat{\mathbf{X}}^\top(\mathbf{G}^{\sharp(ijk)}+\hat{\mathbf{X}}\hat{\mathbf{R}} {\mathbf{Y}}^{\star\top}-\hat{\mathbf{X}}\hat{\mathbf{Y}}^\top)\approx\mathcal{O},\\
	\end{aligned}
	\end{equation}
	
	\noindent where (i) is the result of $d\mathbf{W}^{\sharp(i j k)}=d{\hat{\mathbf{X}}}{\hat{\mathbf{Y}}^{\top}}+\hat{\mathbf{X}}(d\hat{\mathbf{Y}})^\top$. (ii) is approximated by the fact that the rectified local minima ($\hat{\mathbf{X}}\hat{\mathbf{R}}$,$\hat{\mathbf{Y}}\hat{\mathbf{R}}$) are close to the global minima ($\mathbf{X}^{\star}$,$\mathbf{Y}^{\star}$). We manipulate Eq. (\ref{Eq_provePartialZero1}) and Eq. (\ref{Eq_provePartialZero2}) to: 
	\begin{equation}
	\label{Eq_X_Gaussian}
	\begin{aligned}
	&\mathbf{G}^{\sharp(ijk)}\hat{\mathbf{Y}}\approx (\hat{\mathbf{X}}-\mathbf{X}^{\star}\hat{\mathbf{R}}^\top)\hat{\mathbf{Y}}^\top\hat{\mathbf{Y}}\\
	&(\hat{\mathbf{X}}-\mathbf{X}^{\star}\hat{\mathbf{R}}^\top)\approx \mathbf{G}^{\sharp(ijk)}\hat{\mathbf{Y}}(\hat{\mathbf{Y}}^{\top}\hat{\mathbf{Y}})^{-1}\\
	&(\hat{\mathbf{X}}\hat{\mathbf{R}}-\mathbf{X}^{\star})\approx \mathbf{G}^{\sharp(ijk)}\hat{\mathbf{Y}}(\hat{\mathbf{Y}}^{\top}\hat{\mathbf{R}}^\top\hat{\mathbf{R}}\hat{\mathbf{Y}})^{-1}\hat{\mathbf{R}}\\
	&(\hat{\mathbf{X}}\hat{\mathbf{R}}-\mathbf{X}^{\star})\stackrel{(\mathrm{i})}\approx \mathbf{G}^{\sharp(ijk)}\hat{\mathbf{Y}}\hat{\mathbf{R}}(\hat{\mathbf{Y}}^{\top}\hat{\mathbf{R}}^\top\hat{\mathbf{R}}\hat{\mathbf{Y}})^{-1}\\
	&(\hat{\mathbf{R}}^\top\hat{\mathbf{X}}^\top-\mathbf{X}^{\star\top})\mathbf{e}_v\approx [\mathbf{G}^{\sharp(ijk)}\widehat{\mathbf{Y}}(\widehat{\mathbf{Y}}^{\top}\widehat{\mathbf{Y}})^{-1}]^\top\mathbf{e}_v
	\stackrel{(\mathrm{ii})}{\sim} \mathcal{N}(0, \sigma_0^{2}(\Sigma^{\star})^{-1}),
	\end{aligned}
	\end{equation}
	
	\begin{equation}
	\label{Eq_Y_Gaussian}
	\begin{aligned}
	&\hat{\mathbf{X}}^\top\mathbf{G}^{\sharp(ijk)}=\hat{\mathbf{X}}^\top\hat{\mathbf{X}} ({\hat{\mathbf{Y}}}^\top-\hat{\mathbf{R}}{\mathbf{Y}^{\star\top}})\\
	&({\hat{\mathbf{Y}}}^\top-\hat{\mathbf{R}}{\mathbf{Y}^{\star\top}})\approx (\hat{\mathbf{X}}^{\top}{\hat{\mathbf{X}}})^{-1}{\hat{\mathbf{X}}^\top}\mathbf{G}^{\sharp(ijk)}\\
	&(\hat{\mathbf{R}}^\top{\hat{\mathbf{Y}}}^\top-{\mathbf{Y}^{\star\top}})\approx \hat{\mathbf{R}}^\top(\hat{\mathbf{X}}^{\top}\hat{\mathbf{R}}^\top\hat{\mathbf{R}}{\hat{\mathbf{X}}})^{-1}{\hat{\mathbf{X}}^\top}\mathbf{G}^{\sharp(ijk)}\\
	&(\hat{\mathbf{R}}^\top{\hat{\mathbf{Y}}}^\top-{\mathbf{Y}^{\star\top}})\stackrel{(\mathrm{i})}\approx (\hat{\mathbf{X}}^{\top}\hat{\mathbf{R}}^\top\hat{\mathbf{R}}{\hat{\mathbf{X}}})^{-1}\hat{\mathbf{R}}^\top{\hat{\mathbf{X}}^\top}\mathbf{G}^{\sharp(ijk)}\\
	&(\hat{\mathbf{R}}^\top{\hat{\mathbf{Y}}^\top}-{\mathbf{Y}^{\star\top}})\mathbf{e}_v\approx  (\widehat{\mathbf{X}}^{\top}{\widehat{\mathbf{X}}})^{-1}{\widehat{\mathbf{X}}^\top}\mathbf{G}^{\sharp(ijk)} \mathbf{e}_v
	\stackrel{(\mathrm{iii})}{\sim} \mathcal{N}(0, \sigma_0^{2}(\Sigma^{\star})^{-1}),
	\end{aligned}
	\end{equation}
	
	\noindent where the matrix exchange in step (i) is enabled because $(\hat{\mathbf{X}}^{\top}\hat{\mathbf{R}}^\top\hat{\mathbf{R}}{\hat{\mathbf{X}}})$ and $(\hat{\mathbf{Y}}^{\top}\hat{\mathbf{R}}^\top\hat{\mathbf{R}}{\hat{\mathbf{Y}}})$ are diagonal matrices. We denote the rectified local minima ($\widehat{\mathbf{X}}$,$\widehat{\mathbf{Y}}$) equal to ($\hat{\mathbf{X}}\hat{\mathbf{R}}$,$\hat{\mathbf{Y}}\hat{\mathbf{R}}$) for a concise presentation. \par
	
	The proof of step (ii): Each element of the Gaussian noise matrix $\mathbf{G}^{\sharp(ijk)} {\sim} \mathcal{N}(0, \sigma^{2})$. According to Eq. (\ref{Eq_XtX_Sigma}), $\Sigma^{\star}$ is a diagonal matrix and the diagonal element at position $[i,i]$ $(\Sigma^{\star}_{ii})^{-1}$ equals to $1/||\mathbf{X}_{.,i}||_2^2$ or $1/||\mathbf{Y}_{.,i}||_2^2$ where $\widehat{\mathbf{X}}_{.,i}$ or $\widehat{\mathbf{Y}}_{.,i}$ is the $i$th column of matrix $\widehat{\mathbf{X}}$ or $\widehat{\mathbf{Y}}$. \par
	
	The proof of (i): \par 
	\begin{equation}
	\label{Eq_proof_1}
	\begin{aligned}
	&[\mathbf{G}^{\sharp(ijk)}\widehat{\mathbf{Y}}(\widehat{\mathbf{Y}}^{\top}\widehat{\mathbf{Y}})^{-1}]^\top\mathbf{e}_v\\
	&=[\sum_{i=1}^{L}\frac{1}{\|\widehat{\mathbf{Y}}_{.,1}||_2^2}\widehat{\mathbf{Y}}_{i1}\mathbf{G}^{\sharp(ijk)}_{vi} \ \cdots \  \sum_{i=1}^{L}\frac{1}{\|\widehat{\mathbf{Y}}_{,r}||_2^2}\widehat{\mathbf{Y}}_{ir}\mathbf{G}^{\sharp(ijk)}_{vi}].
	\end{aligned}
	\end{equation}
	
	Take the first element $\sum_{i=1}^{L}\frac{1}{\|\widehat{\mathbf{Y}}_{.,1}||_2^2}\widehat{\mathbf{Y}}_{i1}\mathbf{G}^{\sharp(ijk)}_{vi}$ as an example:
	
	\begin{equation}
	\begin{aligned}
	\operatorname{var}(\sum_{i=1}^{L}\frac{1}{\|\widehat{\mathbf{Y}}_{.,1}||_2^2}\widehat{\mathbf{Y}}_{i1}\mathbf{G}^{\sharp(ijk)}_{vi})&=
	(\sum_{i=1}^{L}\frac{1}{\|\widehat{\mathbf{Y}}_{.,1}||_2^2}\widehat{\mathbf{Y}}_{i1})^2\sigma_0^2\\
	&=\frac{1}{(\|\widehat{\mathbf{Y}}_{.,1}||_2^2)^2}(\sum_{i=1}^{L}\widehat{\mathbf{Y}}_{i1})^2\sigma_0^2\\
	&=\frac{1}{(\|\widehat{\mathbf{Y}}_{.,1}||_2^2)}\sigma_0^2\\
	&\approx(\Sigma^{\star}_{11})^{-1}\sigma_0^2.
	\end{aligned}
	\end{equation}
	
	The same conclusion also applies to other elements.
	
	Similarly, the proof of (ii): \par 
	\begin{equation}
	\label{Eq_proof_1}
	\begin{aligned}
	&(\widehat{\mathbf{X}}^{\top}{\widehat{\mathbf{X}}})^{-1}{\widehat{\mathbf{X}}}\mathbf{G}^{\sharp(ijk)}\mathbf{e}_v \\
	&=[\sum_{i=1}^{K}\frac{1}{\|\widehat{\mathbf{X}}_{.,1}||_2^2}\widehat{\mathbf{X}}_{i1}\mathbf{G}^{\sharp(ijk)}_{iv} \ \cdots \  
	\sum_{i=1}^{K}\frac{1}{\|\widehat{\mathbf{X}}_{,r}||_2^2}\widehat{\mathbf{X}}_{i1}\mathbf{G}^{\sharp(ijk)}_{iv}]
	\end{aligned}
	\end{equation}
	
	Take the first element $\sum_{i=1}^{K}\frac{1}{\|\widehat{\mathbf{X}}_{.,1}||_2^2}\widehat{\mathbf{X}}_{iv}\mathbf{G}^{\sharp(ijk)}_{i1}$ as an example:
	
	\begin{equation}
	\begin{aligned}
	\operatorname{var}(\sum_{i=1}^{K}\frac{1}{\|\widehat{\mathbf{X}}_{.,1}||_2^2}\widehat{\mathbf{X}}_{i1}\mathbf{G}^{\sharp(ijk)}_{iv})&=
	(\sum_{i=1}^{K}\frac{1}{\|\widehat{\mathbf{X}}_{.,1}||_2^2}\widehat{\mathbf{X}}_{i1})^2\sigma_0^2\\
	&=\frac{1}{(\|\widehat{\mathbf{X}}_{.,1}||_2^2)^2}(\sum_{i=1}^{K}\widehat{\mathbf{X}}_{i1})^2\sigma_0^2\\
	&=\frac{1}{(\|\widehat{\mathbf{X}}_{.,1}||_2^2)}\sigma_0^2\\
	&\approx(\Sigma^{\star}_{11})^{-1}\sigma_0^2.
	\end{aligned}
	\end{equation}
	
	The same conclusion also applies to other elements. \par
	
	Following Eqs. (\ref{Eq_X_Gaussian}) and (\ref{Eq_Y_Gaussian}), we strictly obtain the conclusion that the residual of Gaussian noise $\mathbf{G}^{(i j k)} \stackrel{\mathrm{i.i.d.}}{\sim} \mathcal{N}(0, \sigma_0^{2})$. Thus, $\Psi_X \in \mathbb{R}^{K \times r}, \Psi_Y \in \mathbb{R}^{L \times r}$ are the negligible residuals due to numerical issues.\par 
	This concludes the proof of $(i)$ and $(ii)$ and also the proof of \textbf{theorem 1}.\par

	\bibliographystyle{IEEEtran}
	\bibliography{reference}
	
	
	\begin{IEEEbiography}
		[{\includegraphics[width=1in,height=1.25in,clip,keepaspectratio]{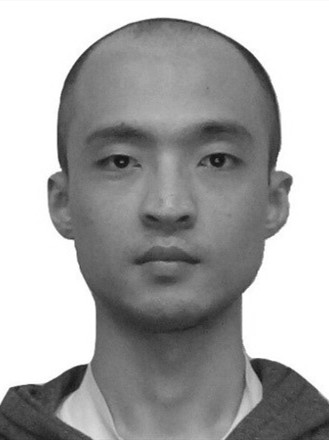}}]{Jingwei Song}
		obtained his doctoral degree in Center of Autonomous System, University of Technology Sydney in 2020. Before that, he received B.E. degree in Remote sensing science and technology from Wuhan University, Wuhan in 2012. He obtained a master's degree in Cartography and Geographic Information System from Institute of Remote Sensing and Digital Earth, Chinese Academy of Sciences in 2015. He is a research associate at the Department of Naval Architecture and Marine Engineering, University of Michigan, Ann Arbor. His research interests include computer vision, simultaneous localization and mapping, and robotics.
	\end{IEEEbiography}
	

	\begin{IEEEbiography}
		[{\includegraphics[width=1in,height=1.25in,clip,keepaspectratio]{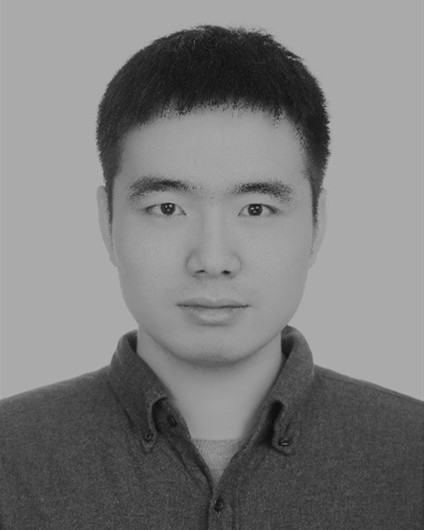}}]{Shaobo Xia}
		received the bachelor’s degree in geodesy and geomatics from the School of Geodesy and Geomatics, Wuhan University, Wuhan, China, in 2013, the master’s degree in cartography and geographic information systems from the Institute of Remote Sensing and Digital Earth, Chinese Academy of Sciences, Beijing, China, in 2016, and the Ph.D. degree in geomatics from the University of Calgary, Calgary, AB, Canada, in 2020. He is an Assistant Professor with the College of Environmental and Resource Sciences, Zhejiang A\&F University, Hangzhou, China. His research interests include mobile mapping and remote sensing.
	\end{IEEEbiography}
	
	\begin{IEEEbiography}
		[{\includegraphics[width=1in,height=1.25in,clip,keepaspectratio]{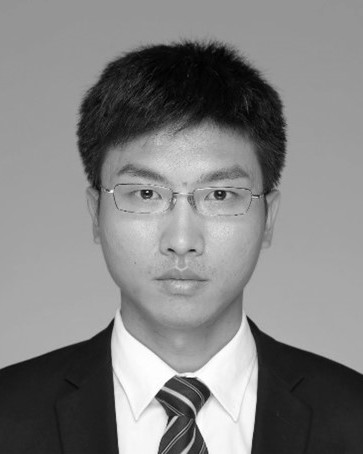}}]{Jun Wang}
		received B.E. degree in Remote sensing science and technology from Wuhan University, Wuhan, China, in 2012 and is currently pursuing the Ph.D. degree in Remote sensing science from University of Chinese Academy of Sciences. He also got a joint Ph.D. program in Robotics from University of Technology, Sydney.  His research interests include computer vision, simultaneous localization and mapping, robotics.
	\end{IEEEbiography}
	
	\begin{IEEEbiography}
		[{\includegraphics[width=1in,height=1.25in,clip,keepaspectratio]{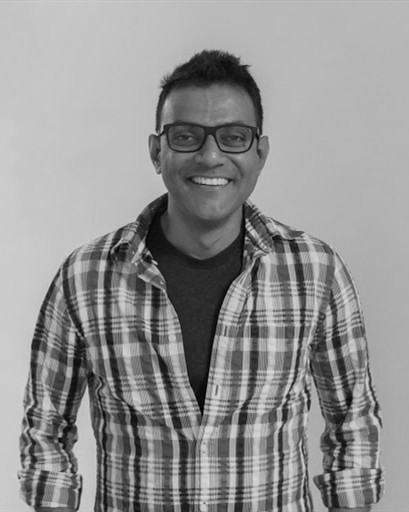}}]{Mitesh Patel}
		received my Ph.D. in Robotics from Center of Autonomous Systems (CAS) at the University of Technology Sydney in 2014 where I focused on modeling a wide spectrum of high-level user activities (also known as activities of daily living) using different probabilistic techniques. Most recently, he was a senior research scientist at FX Palo Alto Laboratory Inc., where he worked on developing indoor localization system using variety of sensors such as RF, RGB, RGB-D images, LiDAR.
	\end{IEEEbiography}

	\begin{IEEEbiography}
		[{\includegraphics[width=1in,height=1.25in,clip,keepaspectratio]{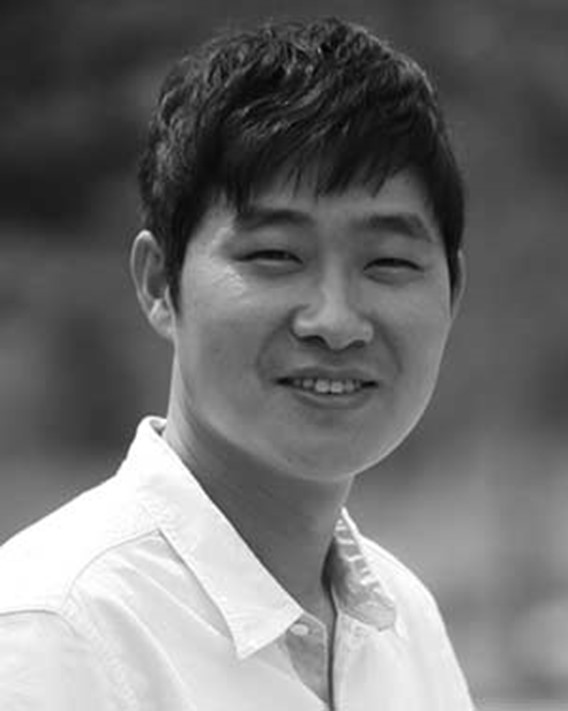}}]{Dong Chen (Member, IEEE)}
		received the bachelor’s degree in computer science from the Qingdao University of Science and Technology, Qingdao, China, in 2005, the master’s degree in cartography and geographical information engineering from the Xi’an University of Science and Technology, Xi’an, China, in 2009, and the Ph.D. degree in geographical information sciences from Beijing Normal University, Beijing, China, in 2013. He is an Associate Professor with Nanjing Forestry University, Nanjing, China. He is also a Post-Doctoral Fellow with the Department of Geomatics Engineering, University of Calgary, Calgary, AB, Canada. His research interests include image-and LiDAR-based segmentation and reconstruction, full-waveform LiDAR data processing, and related remote sensing applications in the field of forest ecosystems.
		
	\end{IEEEbiography}

\end{document}